\documentclass[preprint2]{proto}
\usepackage{natbib}
\bibpunct{(}{)}{;}{a}{,}{,}
\usepackage{amsmath, amsthm, amssymb}
\usepackage{IEEEtrantools}
\usepackage{times}

\usepackage[usenames]{color}

\usepackage{hyperref}      
\hypersetup{backref=true, pagebackref=true, hyperindex=true, breaklinks=true,colorlinks=true,urlcolor=blue, linkcolor=blue,  citecolor=black, pagecolor=red, bookmarks=true, bookmarksopen=true}

\newcommand{\D}{\displaystyle}
\newcommand{\Mstar}{M_*}
\newcommand{\mearth}{M_\oplus}
\newcommand{\mdotst}{\dot{M_*}}
\newcommand{\vect}[1]{\ensuremath{\mbox{\boldmath$#1$}}}
\def \unitvq{\vect}

\begin{document}

\title{\textbf{\LARGE Planet Population Synthesis}}

\author {\textbf{\large Willy Benz}}
\affil{\small\em University of Bern}
\author {\textbf{\large Shigeru Ida}}
\affil{\small\em ELSI, Tokyo Institute of Technology}
\author {\textbf{\large Yann Alibert}}
\affil{\small\em University of Bern}
\author {\textbf{\large Douglas Lin}}
\affil{\small\em University of California at Santa Cruz}
\author {\textbf{\large Christoph Mordasini}}
\affil{\small\em Max Planck Institute for Astronomy}

\begin{abstract}
\baselineskip = 11pt
\leftskip = 0.65in 
\rightskip = 0.65in
\parindent=1pc
{\small With the increasing number of exoplanets discovered, statistical properties of the population as a whole become unique constraints on planet formation models provided a link between the description of the detailed processes playing a role in this formation and the observed population can be established. Planet population synthesis provides such a link. The approach allows to study how different physical models of individual processes (\textit{e.g.,} proto-planetary disc structure and evolution, planetesimal formation, gas accretion, migration, etc.) affect the overall properties of the population of emerging planets. By necessity, planet population synthesis relies on simplified descriptions of complex processes. These descriptions can be obtained from more detailed specialised simulations of these processes. The objective of this chapter is twofold: 1) provide an overview of the physics entering in the two main approaches to planet population synthesis and 2) present some of the results achieved as well as illustrate how it can be used to extract constraints on the models and to help interpret observations.
 \\~\\~\\~}%leave this in to get the correct vertical space after the abstract

\end{abstract}  

\section{\textbf{INTRODUCTION}}
\bigskip

The number of known exoplanets has increased dramatically in recent years (see, e.g., www.exoplanet.eu; \citealt{schneider:2011}). At the time of this writing, over 1000 confirmed exoplanets were known, mostly found through precise radial velocity surveys. Additionally, there are more than 3600 transiting candidate planets found by the {\it Kepler} satellite (see {\it e.g.,} www.kepler.nasa.gov). All these detections have revealed that planets are quite common and that the diversity of existing systems is much larger than was expected from studies our own Solar System. Finally, with increasing numbers of planets, search for correlations and structures in the properties of planets and planetary systems becomes increasingly meaningful. The correlations and structures have pinpointed the importance of complex interaction processes taking place during the formation stages of the planets (e.g. planetary migration).

These insights were essentially gained by the fact that, for the first time, a  large set of planets was available to study and statistical analysis became possible. The analysis of the characteristics of an ensemble of objects as well as of the differences between individual objects is a standard approach in astrophysics and has been applied successfully in a number of areas (\textit{e.g.,} galactic evolution). 

Planet population synthesis in the context of the core accretion scenario has been pioneered by \citet{idalin:2004a} in an effort to develop a deterministic model of planetary formation allowing a direct comparison with the observed population of exoplanets. They presented formation models for planets orbiting solar-type stars but neglected the effect of type I migration on the basis that its efficiency was poorly determined. They simulated the mass-semi-major axis distribution of planets for stars of different metallicity and masses \citep{idalin:2004b, idalin:2005}. \citet{burkertida:2007} applied this model to discuss a potential period gap in the observed gas giant distribution orbiting stars more massive than the sun. \citet{currie:2009} discussed the same issue with a similar model. \citet{paynelodato:2007} discussed planets orbiting brown dwarfs using the model by \citet{idalin:2004b}.

In \citet{idalin:2008a}, type I migration was incorporated, using a conventional isothermal formula \citep[e.g.][]{tanakaetal:2002} with an efficiency factor that uniformly decreases the migration speed, because the predicted formation efficiency of gas giants with the full strength of the migration is too low to be consistent with observations and to allow for uncertainties in the theoretically derived migration speed. \citet{idalin:2008b} considered the effect of a potential migration trap due to a snow line. With a similar model, \citet{migueletal:2011a, migueletal:2011b} studied a dependence on initial disc models. \citet{mordasinietal:2009a} developed a model based on more detailed calculation of gas envelope contraction, disc evolution and planetesimal dynamics. Type I migration was treated in a similar way to \citet{idalin:2008a}.  They applied their model   \citep{mordasinietal:2009b} for statistical comparisons with the then known population of giant extrasolar planets.  \citet{alibertetal:2011} studied the impact of the stellar mass on planetary populations while \citet{mordasinietal:2012c} investigated how important properties of the proto-planetary disc (mass, metallicity, lifetime) translate into planetary properties. In an attempt to further couple formation models to the major observable physical characteristics of a planet (besides mass and semi-major axis also radius, luminosity and bulk composition) \citet{mordasinietal:2012} added  to the model the long-term evolution of the planets after formation (cooling and contraction). This enabled statistical comparisons with results of transit and (in future) direct imaging surveys, and in particular  the result of the Kepler satellite \citep{mordasinietal:2012b}.

Since planet-planet scattering is a chaotic process, including such effects in an otherwise deterministic calculations was not easy. \citet{alibertetal:2013} incorporated a full N-body integrator with collision detections  in order to simulate planet-planet interactions. While this approach adds a significant computational burden, it has the advantage to handle all dynamical aspects (including resonances) correctly. (Note that \citet{bromleykenyon:2006, bromleykenyon:2011, thommesetal:2008, hellarynelson:2012} also developed hybrid N-body simulations, although they did not present much statistical discussions of predicted planet distributions). \citet{idalin:2010} and  \citet{idaetal:2013} took another approach in which the planet-planet interactions (scatterings, ejections, collisions) are treated in a Monte Carlo fashion calibrated by numerical simulations.  Although relatively complicated multiple steps are needed for the Monte Carlo method to reproduce accurate enough predictions for the statistical purposes of population synthesis, it is much faster than direct N-body simulations.

\bigskip
\section{\textbf{THE PHYSICS OF POPULATION SYNTHESIS}}
\label{sec:physics}
\bigskip

As its name indicates, the goal of planet population synthesis is to allow the computing of a full planet population given a suitable set of initial conditions. Practically, this requires a full planet formation model that computes the final characteristics of planets from specific initial conditions. The physics behind the formation model will be discussed in this section while matters related to initial conditions are presented in section \ref{sec:ic}. 

By nature, an end-to-end simulation of the formation of even a single planet is probably impossible to carry out in a complete and detailed manner. Hence, assumptions have to be made in order to make the problem tractable. In this approach the difficulty is to identify how far the problem can be simplified while still conserving the overall properties of the emerging planet population, including their mean values and dispersions. Guidance must come as much as possible from observations and from detailed modelling of all the individual processes entering in the computations of a planet population. 

\begin{figure*}
  \epsscale{1.5}
  \plotone{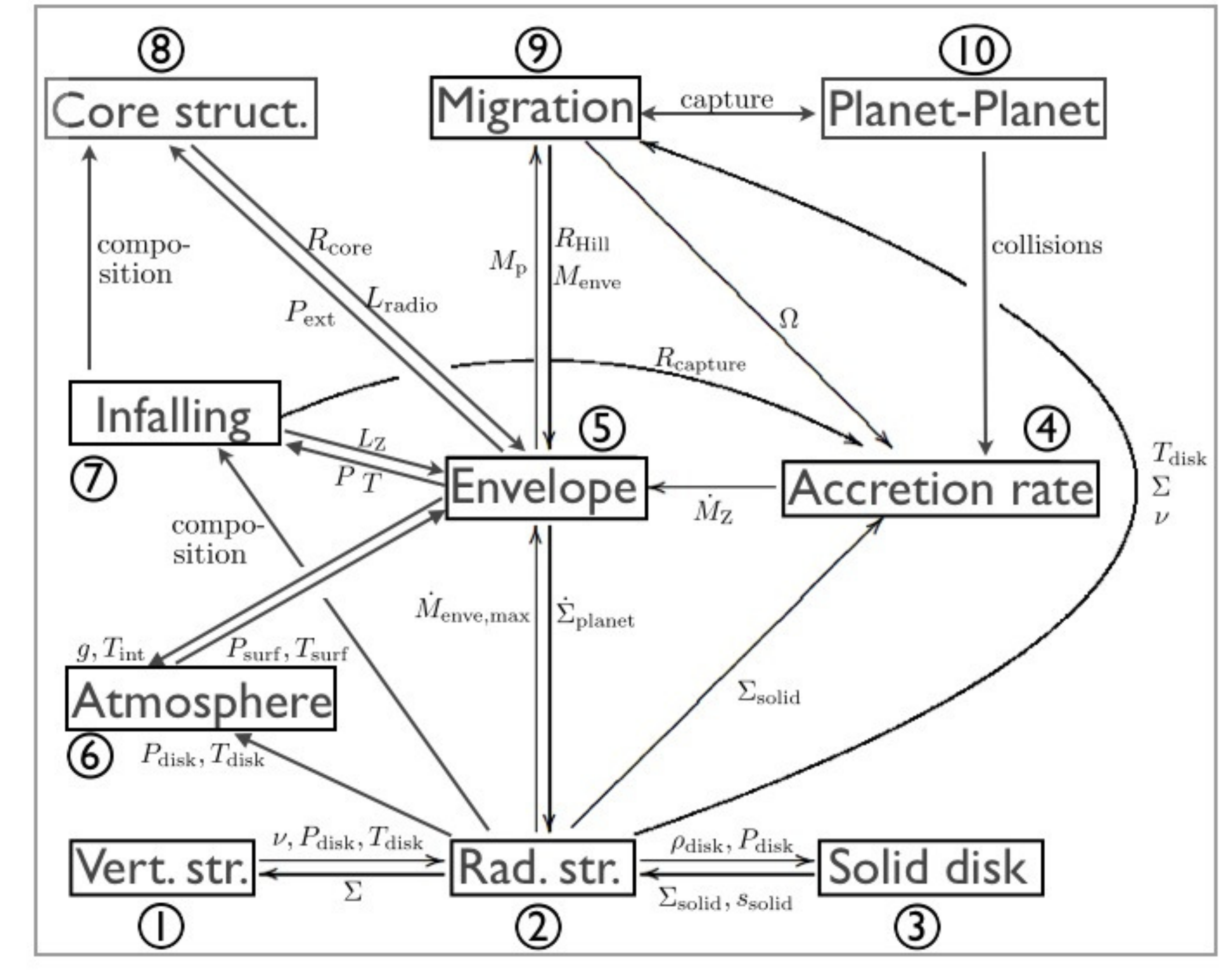}
  \caption{\small Schematic of the coupling between the different processes entering in the computation of a self-consistent planet formation model. Quantities exchanged by the different modules are indicated along arrows.}   
  \label{fig:schematic}
\end{figure*}
Figure~\ref{fig:schematic} provides an overview of all the elements that enter in a self-consistent planet formation model. As can be seen from this figure, a large number of processes enter in the physical computation of the formation of a planet. By necessity, each process has to be described in rather simplified physical terms. We stress that most of these descriptions are not specific to population synthesis but are commonly used throughout the literature to discuss these individual processes. The essence of population synthesis consists therefore of coupling these processes in a physically meaningful and consistent manner. This is especially important as many of the processes have comparable timescales. 
 
In what follows we describe the essential ingredients of the planet formation model as they have been worked out in a series of papers by Ida and Lin \citep[][hereafter referred to as "IL"]{idalin:2004a, idalin:2004b, idalin:2005, idalin:2008a, idalin:2008b, idalin:2010, idaetal:2013} and by a series of papers by Alibert, Mordasini, and Benz \citep[][hereafter referred to as "AMB"]{mordasinietal:2009a, mordasinietal:2009b, alibertetal:2011, mordasinietal:2012b,mordasinietal:2012c,alibertetal:2013}  with significant contributions by various collaborators at the University of Bern (A. Fortier in particular) and at the Max Planck Institute for Astronomy. We opted for this relatively detailed approach as we believe that the extent of the physical description of the processes entering in these models is not always fully appreciated as it is disseminated  throughout a number of papers.

\subsection{Structure and evolution of the proto-planetary disc}
\label{subsec:gasdiscstruc}
%------------------------------------------------------------------------------------

\noindent Capturing the structure and evolution of the proto-planetary disc is important since the migration rate of planets, as well as their internal structure (through the planetary surface conditions), and the amount of gas they can accrete is determined, at least partially, by the disc (module 1 and 2 in Fig.~\ref{fig:schematic}). Different level of complexity can be used to describe the disc.  Existing state-of-the-art fully 3D self-consistent magneto-hydrodynamical models are not applicable for population synthesis as the computational resources involved are such that only relatively short timespans can be modelled. Given both observational incompleteness and theoretical uncertainties in the detailed structure of discs, one simple approach consists of assuming that the gas surface density follows exponential decay with a characteristic disc evolution timescale (comparable to observed disc lifetimes, \textit{i.e.}, of a few million years) \citep[e.g.][]{idalin:2004a}. This approach has the advantage of being extremely fast and to allow exploration of a large number of models.  However, this approach does not provide self-consistently a relation between the disc surface density and the local pressure and temperature which enter in the calculation of the structure of the growing planets as well as in their migration rates. An intermediate approach consists of using a model of viscously evolving discs \citep{shakurasunyaev:1973} for which the local vertical structure can be computed \citep{papaloizouterquem:1999, alibertetal:2005a}. Such an intermediate approach has the advantage of providing the full structure of the disc and to provide the framework to include self-consistently additional physical processes such as photo-evaporation from the central star and/or nearby stars, the existence of dead-zones, and irradiation from the central star. Both approaches are briefly outlined below.

IL adopt the minimum mass solar nebula (MMSN) model \citep{hayashi:1981} as a fiducial set of initial conditions and introduce multiplicative factors ($f_{d}$ and $f_g$) to scale the MMSN disc surface densities of gas ($\Sigma$) and planetesimals ($\Sigma_s$). IL  set 
\begin{equation} 
   \Sigma = \Sigma_{10} f_{g} (r/ {\rm 10 AU})^{-q},
   \label{eq:sigma_gas}
\end{equation}
where a normalization factor $\Sigma_{10} = 75 {\rm g/cm}^2$  corresponds to 1.4 times of $\Sigma$ at 10AU of the MMSN model. The inner disc boundary where $\Sigma$ vanishes is set at $\sim 0.04$AU. IL often use a power law exponent $q = 1$ which corresponds to the self-similar steady accretion disc model with a constant $\alpha$ viscosity, rather than the original MMSN model for which $q = 1.5$. However, they found that this does not considerably affect the results. 

Neglecting the detailed energy balance in the disc, IL adopt the equilibrium temperature distribution of optically thin discs prescribed by \citet{hayashi:1981},
\begin{equation}
   T = 280 \left(\frac{r}{1{\rm AU}}\right)^{-1/2}
    \left(\frac{L_*}{L_{\odot}}\right)^{1/4} {\rm K},
   \label{eq:temp_dist}
\end{equation}
where $L_*$ and $L_{\odot}$ are respectively stellar and solar luminosity. IL determine the position of the ice line ($a_{\rm ice}$) as the location at which $T=170$K, which translates for a optically thin disc into (Eq.~[\ref{eq:temp_dist}])
\begin{equation}
   a_{\rm ice} = 2.7 (L_\ast/L_\odot)^{1/2} {\rm AU}.
   \label{eq:a_ice}
\end{equation}

Due to viscous diffusion and photo-evaporation, $f_{g}$ decreases with time. For simplicity, IL adopt 
\begin{equation}
   f_{g} = f_{g,0} \exp(-t/\tau_{\rm dep}),
   \label{eq:gas_exp_decay}
\end{equation}
where $\tau_{\rm dep}$ is the disc lifetime (for detailed discussion, see \citealt{idalin:2008a}).  IL use $\tau_{\rm dep}$ as a free parameter ranging from $10^6$yrs to $10^7$yrs. The self-similar solution with $\Sigma \propto r^{-1}$ has an asymptotic exponential cut-off at radius $r_{\rm m}$ of the maximum viscous couple. In the region at $r < r_{\rm m}$, $\Sigma$ decreases uniformly independent of $r$ as the exponential decay does, although the time dependence is slightly different. Note that this treatment is relevant in the regime where the disc mass depletion rate by photo-evaporation is so low that it does not affect disc evolution by viscous diffusion until the last phase of disc depletion. 

The disc structure and evolution in the AMB model does not assume any power law structure but is calculated from assuming local hydrostatic equilibrium. In this case, the vertical structure of the disc can be computed from
\begin{equation}
   \frac{1}{\rho} \frac{\partial P}{\partial z} = - \Omega^2 z
\label{eq_disk_hydro}
,
\end{equation}
where $z$ is the vertical coordinate, $\rho$ the density and $P$ the pressure, and $\Omega$ is angular frequency of the disc. The disc is assumed to be Keplerian, therefore $\Omega^2 = G \Mstar / r^3$, $G$ being the gravitational constant and $\Mstar$ the mass of the central star. This equation is solved together with the energy equation which states that the energy produced by viscosity is removed by the radiative flux:
\begin{equation}
   \frac{\partial F}{\partial z} = \frac{9}{4} \rho \nu \Omega^2
\label{eq_disk_ener}
,
\end{equation}
where $F$ is the radiative flux  \citep[see also][]{rudenlin:1986}. Assuming an optically thick medium, the radiative flux is written
\begin{equation}
    F = - \frac{16 \pi \sigma T^3}{3 \kappa \rho} \frac{\partial T}{\partial z}
\label{eq_rad_flux}
,
\end{equation}
where $T$ is the temperature, $\kappa$ is the opacity, and $\sigma$ is the Stefan-Boltzmann constant.  The viscosity is calculated using the standard $\alpha-$parametrization $\nu = \alpha c_s^2 / \Omega$ where the speed of sound $c_s^2$ is determined from the equation of state. This set of three differential equations can be solved with the addition of suitable boundary conditions \citep{papaloizouterquem:1999}:
\begin{equation}
   P_s = \frac{\Omega^2 H \tau_{\rm ab} }{\kappa_s}
,
\end{equation}
\begin{equation}
   F_s= \frac{3}{8 \pi} \mdotst \Omega^2
,
\end{equation}
\begin{equation}
   2 \sigma (T_s^4 - T_b^4) - \frac{\alpha k T_s \Omega}{8 \mu m_H \kappa_s} - \frac{3}{8\pi} \mdotst \Omega^2 = 0
,
\end{equation}
\begin{equation}
   F(z=0) = 0
.
\end{equation}
where the subscript $s$ refers to values taken at $z=H$, $\tau_{\rm ab}$ is the optical depth between the surface of the disc ($z = H$) and infinity, $T_b$ is the background temperature, $k$ is the Boltzmann constant, $\mu$ the mean molecular mass of the gas,  and $m_H$ the mass of the hydrogen atom. In the above equations, $\mdotst$ is the equilibrium accretion rate defined by $\mdotst \equiv 3 \pi \tilde{\nu} \Sigma$ where $\Sigma \equiv \int_{-H}^{H} \rho dz$ is the surface density, and $\tilde{\nu}$ the effective viscosity defined by
\begin{equation}
   \tilde{\nu} \equiv \frac{\int_{-H}^{H} \nu \rho dz}{\Sigma}
   \label{nu_eff}
   .
\end{equation}

The effect of the irradiation of the central star can, in its simplest form, be included by suitably modifying the surface temperature of the disc \citep{fouchetetal:2012} in the form
\begin{equation}
   T_s^4 = T_{s, noirr}^4 + T_{s,irr}^4
,
\end{equation}
where $T_{s, noirr}$ is the temperature due to viscous heating given in the above and the irradiation temperature is given by \citet{huesoguillot:2005}
 \begin{IEEEeqnarray}{rCl}
       T_{s,irr} & = & T_* \left[\frac{2}{3\pi}\left(\frac{R_*}{r}\right)^3 +
       			   \frac{1}{2} \left(\frac{R_*}{r}\right)^2 \left(\frac{H_P}{r}\right) \right. \nonumber \\
			        && \left. \left(\frac{d \ln H_p}{d \ln r} -1\right) \right]^{1/4}
			   ,
\end{IEEEeqnarray}
where $R_*$ is the stellar radius, $r$ is the distance to the star and $H_P$ is the pressure scale height defined as $\rho(z = H_P) = e^{-1/2} \rho(z=0)$ \cite[see also][]{ChiangGoldreich:1997, GaraudLin:2007}. 

The radial evolution of the disc (module 2 in Fig.~\ref{fig:schematic}) is provided by the standard viscous disc evolution equation \citep{lyndenbellpringle:1974} complemented by appropriate terms describing the gas accreted by the planets and the one lost by photo-evaporation. It is customary to write this equation in terms of the evolution of the surface density $\Sigma(r)$

\begin{equation}
   \frac{\partial \Sigma}{\partial t} = \frac{3}{r} \frac{\partial}{\partial r}\left[r^{1/2} \frac{\partial}{\partial r}(\tilde{\nu} \Sigma r^{1/2}) \right] + \dot{\Sigma}_{\rm w}(r)
   + \dot{Q}_{\rm planet}
   \label{disk_surf_evol}
,
\end{equation}
where $\dot{\Sigma}_w (r)$ describes the sink term associated with photo-evaporation caused by the host star itself (internal photo-evaporation) as well as by close-by massive stars (external photo-evaporation) \citep[for a detailed discussion, see][]{mordasinietal:2012b} and $\dot{Q}_{\rm planet}$ represents the rate at which gas is being accreted by the growing planets.

Internal photo-evaporation due to the stellar EUV radiation is modelled according to \citet{clarkeetal:2001} based on the ``weak stellar wind'' scenario of  \citet{hollenbachetal:1994}. It leads to mass loss concentrated on an annulus around $\beta_{\rm II}  R_{g,{\rm II}}$ where $R_{g,{\rm II}}\approx 7$ AU for a 1 $M_{\odot}$ star is the gravitational radius for ionized hydrogen (mass $m_{\rm H}$) and a speed of sound $c_{s,{\rm II}}$ associated with a temperature of approximately $10^{4}$ K, and $\beta_{\rm II}$ is a parameter reflecting that some mass loss occurs already inside of  $R_{g,{\rm II}}$.  The decay rate of disc surface density due to the mass loss is
\begin{equation}
\dot{\Sigma}_{w,{\rm int}}  =\begin{cases}
0 & \textrm{for}\  r < \beta_{\rm II}  R_{g,{\rm II}}\\
2 c_{s,{\rm II}} n_0(r) m_{\rm H} & \textrm{otherwise}.
\end{cases}
\end{equation}
The density of ions at the base of the wind $n_0(r)$ as a function of distance is approximately
\begin{equation}
n_{0}(r)=n_{0}(R_{\rm 14}) \left(\frac{r}{\beta_{\rm II}  R_{g,{\rm II}}}\right)^{-5/2}.
\end{equation}
The density  $n_{0}$ at a normalization radius $R_{\rm 14}$ is given by radiation-hydrodynamic simulations \citep{hollenbachetal:1994} and depends on the ionizing photon luminosity of the central star.

For FUV-driven external photo-evaporation, the mass loss  outside of a critical radius $\beta_{\rm I} R_{g,{\rm I}}$ can be written as \citep{matsuyamaetal:2003}:
\begin{equation}
   \dot{\Sigma}_{\rm w,ext} = 
   \begin{cases} 
     	0, &  r \le \beta_{\rm I} R_{g,{\rm I}}  \\
      \frac{\D\dot{M}_{\rm wind,ext}}{\D \pi ( R_{\rm max}^{2}-\beta_{\rm I}^{2} R_{g,{\rm I}}^{2})}, &  r >  \beta_{\rm I} R_{g,{\rm I}}
   \end{cases}
   \label{eq:BAMextphotoevap}
\end{equation}
where $R_{g,{\rm I}}$ is the gravitational radius for  neutral hydrogen at a temperature of approximately 1000 K ($\approx 140$ AU for a $1 M_{\odot}$ star), while $R_{\rm max}$ is a fixed outer radius. The total mass lost through photo-evaporation is a free parameter and is set by $\dot{M}_{\rm wind,ext}$ in such a way that, together with the viscous evolution, the distribution of disc lifetimes is in agreement with observations. 

The term $\dot{Q}_{\rm planet}$ is obtained from the computation of the amount of gas accreted by the planet(s) as described in Sect. \ref{subsec:gasaccretion} . The amount accreted is removed from disc over an annulus centred on the planet(s), with a width equal to the corresponding Hill radius
\begin{equation}
   R_H = a \left( \frac{M}{3M_*}\right)^{1/3}
\end{equation}
where $M$ is the mass of the planet, $a$ its semi-major axis, and $M_*$ the mass of the central star.

\subsection{Structure and evolution of the disc of planetesimals}
\label{structplandisc}
%-------------------------------------------------------------------------------------

\noindent  The structure and dynamical evolution of the disc of planetesimals (module 3 in Fig.~\ref{fig:schematic}) is essential in order to capture the essence of the growth of the planets. Both IL and AMB have considered so far only two types of planetesimals: rocky and icy. A planetesimals is declared rocky or icy according to its initial position in the disc and does not change nature subsequently. Planetesimals located inside the ice line are rocky while those locate outside are icy. The ice line ($a_{\rm ice}$) is defined as the distance from the star where the temperature of the gas drops below the ice condensation temperature of approximately 170K (IL) and $160$ K (AMB) for the typical pressure range encountered (see section \ref{subsec:gasdiscstruc} for how this temperature is computed).  Note that while "icy" planetesimals means planetesimals composed mainly of ice, they also have a rocky component. Following \citet{hayashi:1981}, IL and AMB assume the mass fraction of the rocky component in icy planetesimals to be $1/4.2$ and $1/4$, respectively.

As far as the radial distribution of solids is concerned, IL set the surface density of planetesimals ($\Sigma_s$) 
\begin{equation}
   \Sigma_s = \Sigma_{s,10} \eta_{\rm ice} f_d (r/ {\rm 10 AU})^{-q_s}, 
   \label{eq:sigma_dust} 
\end{equation}
where a normalization factor $\Sigma_{s,10} = 0.32 {\rm g/cm}^2$ corresponds to 1.4 times of $\Sigma_s$ at 10AU of the MMSN model, and the step
function $\eta_{\rm ice} = 1$ inside the ice line at $a_{\rm ice}$ (Eq.~[\ref{eq:a_ice}]) and 4.2 for $r > a_{\rm ice}$. IL usually adopt $q_s = 1.5$ according to MMSN,  because dust grains suffer inward migration due to gas drag and can be more concentrated in the inner regions than gas components. But, $q_s$ can be similar to the exponent of radial distribution of gas ($\sim 1$). The initial disc metallicity [Fe/H] is used to relate $f_{d,0}$ and $f_{g,0}$ as $f_{d,0}=f_{g,0} 10^{{\rm [Fe/H]}}$, where $f_{d,0}$ and $f_{g,0}$ are initial values of $f_{d}$ and $f_{g}$, respectively.

AMB set the initial surface density of planetesimals proportional to the gas surface density:
\begin{equation}
   \Sigma_s (r,t=0)= f_{D/G} f_{R/I}(r) \Sigma(r, t=0)
   \label{eq:sigmasolidBAM}
,
\end{equation}
where $f_{D/G}$ is the dust-to-gas ratio, which is equal to $f_{d,0}/f_{g,0}$ in the IL expression, $f_{R/I}$ the rock-to-ice ratio (\textit{i.e.}, a factor taking into account the degree of condensation of the volatiles), which has the same role as $\eta_{ice}$ in the IL expression.

The above equation provides the spatial distribution in terms of surface density of the planetesimals. In addition, the dynamics of these planetesimals needs to be specified as the motion of the planetesimals will ultimately determine their collision rate and the growth rate of planets. The dynamics of the planetesimals is determined by three different processes: 1) nebular gas drag, 2) gravitational stirring by growing proto-planets  also known as viscous stirring, and 3) mutual gravitational interactions between the planetesimals.

IL directly give a proto-planet's growth timescale ($\tau_{\rm c,acc}$) as a function of its mass ($M_{\rm c}$) and semi-major axis and disc surface density ($f_d$ and $f_g$). The expression for the growth timescale used by IL is provided in section \ref{subsec:solidacc}. Because runaway growth is quickly transformed into oligarchic growth \citep{idamakino:1993}, the system is reduced to a bimodal population of proto-planets and small planetesimals \citep{kokuboida:1998, kokuboida:2002}. Since the stirring of planetesimal velocity dispersion by the proto-planets is dominating over the mutual stirring between the planetesimals, IL neglected the latter effect. In this approach, the growth rate of a proto-planet is determined by spatial mass density of the planetesimals and relative velocity between a proto-planet and the planetesimals. Dynamical friction from the small planetesimals damps velocity dispersion (eccentricity and inclination) of the proto-planets below those of the planetesimals \citep{idamakino:1992}, so that the relative velocity is dominated by the velocity dispersion of the planetesimals. The velocity dispersion of the planetesimals is determined by a balance between gravitational stirring by a nearby proto-planet and damping due to gas drag.  

In contrast, AMB solve much more detailed equations of evolution of eccentricity $e$ and inclination $i$ of the planetesimals instead of using a final formula derived from simulations. In this approach, the evolution of these quantities is determined by computing explicitly the contribution from all three processes mentioned above \citep{fortieretal:2013}.

\begin{equation} 
   \frac{de^2}{dt}  = \left( \frac{de^2}{dt} \right)_{drag} + \left( \frac{de^2}{dt} \right)_{VS,M} + \left( \frac{de^2}{dt} \right)_{VS,m}
\end{equation}
\begin{equation} 
   \frac{di^2}{dt}  = \left( \frac{di^2}{dt} \right)_{drag} + \left( \frac{di^2}{dt} \right)_{VS,M} + \left( \frac{di^2}{dt} \right)_{VS,m},
\end{equation}
where the first terms are damping due to gas drag, the second terms are excitations due to scattering by proto-planets, and the third terms are those by the other planetesimals. No hypothesis is made as far as an equilibrium between any two terms is concerned in solving the above equations.

\subsection{Accretion of solids}
\label{subsec:solidacc}
%-----------------------------------------

\noindent  A proto-planet grows in mass by accreting planetesimals (module 4 in Fig.~\ref{fig:schematic}) and nebular gas. Since proto-planets are seeded by very small masses (typically with a mass of $10^{20}$g$\sim 10^{-8} \mearth$ for IL and $10^{-2} \mearth$ for AMB; note however that the initial mass does not affect the results since planetesimal accretion is slower in later phase), they are growing initially essentially through the accretion of planetesimals. Later, as the proto-planets reach larger masses, they will be able to grow a gaseous envelope by gravitationally binding nebular gas. 

In the IL approach, the growth of the proto-planet is specified by a growth timescale that is defined by its mass ($M_c$) and semi-major axis ($a$) and the local surface density. This timescale can be computed \citep{idalin:2004a,idalin:2010} from the analytically evaluated relative velocity based on planetesimal dynamics revealed by N-body simulations \citep{kokuboida:1998, kokuboida:2002} and Monte Carlo 3-body simulations \citep{ohtsukietal:2002} and gas drag laws derived by \citet{adachietal:1976}:
\begin{IEEEeqnarray}{rCl}
   \tau_{\rm c,acc} & = & 3.5 \times 10^{5} \eta_{\rm ice}^{-1} f_d^{-1} f_{\rm g}^{-2/5} 
    \left( \frac{a}{1{\rm AU}} \right)^{5/2} \nonumber \\
    &&  \times \left(\frac{M_{\rm c}}{M_{\oplus}} \right)^{1/3} \left(\frac{M_\ast}{M_{\odot}} \right)^{-1/6} {\rm yrs},
   \label{eq:m_grow0}
\end{IEEEeqnarray}
where the mass of the typical field planetesimals is set to be $m=10^{20}$g.  

In the AMB approach, the accretion rate of solid by a core of mass $M_{core}$ is explicitly computed (\citealt{chambers:2006}; \citealt{fortieretal:2013}),
\begin{equation}
   \frac{dM_{core}}{dt} = \left( \frac{2 \pi \Sigma_s R_H^2}{P_{orb}} \right) P_{coll}
   \label{eq:solacc}
\end{equation}
where $\Sigma_s$  is the surface density of planetesimals at the planet location and $P_{orb}$ the orbital period of the planet. The collision probability $P_{coll}$ for a planet to accrete planetesimals depends upon two important parameters: 1) the relative velocities between planets and planetesimals and 2) the presence of an atmosphere large enough to dissipate enough of the kinetic energy of the planetesimals to result in merging. The relative velocities between planetesimals and proto-planets as a function of time can be derived from the knowledge of the time evolution of their respective eccentricities and inclinations (see section \ref{structplandisc}). 

When the planetary core reaches a mass larger than $\sim 1 \mearth$, its gas envelope becomes massive enough to affect the dynamics of planetesimals that penetrate it. As a result of gas drag, the effective cross section of the planet is increased. Such an effect must be computed  module 7 in Fig.~\ref{fig:schematic} by numerically solving for the motion of planetesimals in the planetary envelope under the effects of gravity, gas drag, thermal ablation, and mechanical disruption \citep{podolaketal:1988,alibertetal:2005a}. Alternatively, it is possible to use fits of similar calculations that provide directly the planet cross section \citep[e.g.][]{inabaetal:2001}. The same module also calculates the mass (and energy) deposition of the planetesimals in the proto-planetary envelope \citep{mordasinietal:2006}, yielding the heavy element enrichment of the envelope. Observationally, this is of interest for transit and spectroscopic studies of  extrasolar planets \citep[e.g.][]{fortneyetal:2013}. 

The relations used so far by both IL and AMB are derived from theoretical considerations and comparisons with N-body simulations assuming a single isolated proto-planet embedded in a large number of smaller planetesimals. However, during the formation of planetary systems, several proto-planets grow concurrently and some times sufficiently close to each other for the feeding zone to overlap. In this case, two or more proto-planets compete for the available planetesimals. N-body simulations of such a situation \citep{alibertetal:2013} have shown that the proto-planets are so efficient in scattering the planetesimals that the latter are homogenised over the sum of the feeding zone of the neighbouring growing proto-planets'. This, in turn, changes the mass reservoir of solids available to accrete from and therefore changes the growth of the proto-planets. Finally, gas drag combined proto-planets' tidal perturbation may lead to the clearing of planetesimal gaps which may also reduce $d M_{core}/dt$ \citep{zhoulin:2007}.

\subsection{Envelope structure and accretion of gas}
\label{subsec:gasaccretion}
%----------------------------------------------------------------------

\noindent  The knowledge of the planetary interior structure is essential to compute the accretion rate of gas (module 5 and 6  in Fig.~\ref{fig:schematic}). Indeed, gas accretion depends crucially on the ability of a planet to cool and radiate away the energy gained by the accreting of planetesimals. This is nicely exemplified by the dependence of planetary internal structure on the typical opacity in the envelope (see section \ref{sec:results}). 

IL use a fitting formula for the critical core mass ($M_{\rm c,hydro} $) beyond which atmospheric pressure no longer supports gas envelope against the planetary gravity (\textit{i.e.}, no hydrostatic equilibrium exists) as well as to describe the  quasi-static envelope contraction afterward. These formulas were obtained by detailed 1D calculations of envelope structure and radiative/convective heat transfer as described below.

\citet{ikomaetal:2000} carried out a 1D calculation of envelope structure similar to \citet{bodenheimerpollack:1986} (also see below) with a broad range of parameters and derived the critical core mass as
\begin{equation}
   M_{\rm c,hydro} \simeq 10 \left( \frac{\dot{M}_{\rm c}}{10^{-6}M_{\oplus}/ {\rm yr}}\right)^{0.25} M_{\oplus},
   \label{eq:crit_core_mass}
\end{equation}
where the dependence on the opacity in the envelope \citep[e.g.,][]{idalin:2004a,horiikoma:2010} is neglected, because opacity in the envelope is highly uncertain. Note that $M_{\rm c,hydro}$ depends on the planetesimal-accretion rate $\dot M_{\rm c}$. Since $M_{\rm c,hydro}$ can be comparable to an Earth-mass $M_\oplus$ after the core accretes most of planetesimals in its feeding zone, whether the core becomes a gas giant planet is actually regulated by a timescale of the subsequent quasi-static envelope contraction (Kelvin-Helmholtz contraction time) rather than the value of $M_{\rm c,hydro}$.

Because the contraction of the gas envelope also releases energy to produce pressure to support the gas envelope itself, the contraction is quasi-static. Its rate is still regulated by the efficiency of radiative/convective transfer in the envelope such that
\begin{equation}
   \frac{dM_{\rm planet}}{dt} \simeq \frac{M_{\rm planet}}{\tau_{\rm KH}},
   \label{eq:mgsdot}
\end{equation}
where $M_{\rm planet}$ is the planet mass including gas envelope. Based on the results by 1D calculations \citep{ikomaetal:2001}, IL approximate the Kelvin-Helmholtz contraction timescale $\tau_{\rm KH}$ of the envelope with
\begin{equation}
   \tau_{\rm KH} \simeq \tau_{\rm KH1}
   \left(\frac{M_{\rm planet}}{M_{\oplus}}\right)^{-k2},
   \label{eq:tau_KH}
\end{equation}
where $\tau_{\rm KH1}$ is the contraction timescale for $M_{\rm planet}=M_{\oplus}$. Since there are uncertainties associated with dust sedimentation and  opacity in the envelope \citep{pollacketal:1996, helledetal:2008, horiikoma:2011}, IL adopt a range of values $\tau_{\rm KH1} = 10^8-10^{10}$ years and $k2 = 3$--4 with nominal parameters of $k2 = 3$ and $\tau_{\rm KH1} = 10^9$ years. Eq.~(\ref{eq:mgsdot}) shows that $dM_{\rm planet}/dt$ rapidly increases  as $M_{\rm planet}$ grows.  However, it is limited by the global gas accretion rate throughout the disc and by the process of gap formation near the proto-planets' orbits, as discussed later.

The AMB approach consists of solving the standard internal structure equations \citep{bodenheimerpollack:1986} 
\begin{equation}\label{eq:masscons}
   \frac{dr}{dM_r}= \frac{1}{4 \pi \rho r^2}
\end{equation}
\begin{equation}\label{eq:hydrostaticequilibrium}
   \frac{dP}{dM_r}= - \frac{ G M_r}{4 \pi r^4}
\end{equation}
\begin{equation}
   \frac{dT}{dP}= \nabla_{ad}\    \textrm{or}\   \nabla_{rad}
\end{equation}
where $r, P, T$ are the radius, pressure, and temperature which are specified as a function of the mass $M_r$ which represents the mass inside a sphere of radius $r$  (including the mass of the core  $M_{\rm core}$). Stability against convection is checked using the Schwarzschild criterion {\citep[e.g.][]{kippenhahnweigert:1994}. Depending upon if convection is present or not, the adiabatic gradient ($\nabla_{ad}$) or the radiative gradient ($\nabla_{rad}$) is used.  These equations are solved together with the equation of state (EOS) by \citet{saumonetal:1995}. The opacity is taken from \citet{belllin:1994}. \citet{podolak:2003} and \citet{movshovitzpodolak:2008} have argued that grain opacities are significantly reduced in planetary envelopes as compared to the interstellar medium. Reducing the grain opacity allows runaway accretion to occur at smaller core masses and therefore speeds-up the giant formation timescale (\citealt{pollacketal:1996}; \citealt{hubickyjetal:2005}).

In order to gain computing time and to avoid numerical convergence difficulties, the energy equation is not solved.  Instead the procedure outlined by \citet{mordasinietal:2012}, based on total energy conservation, is adopted with a small improvement which allows to take the energy of the core into account as well as described in \citep{fortieretal:2013}. The total luminosity $L=L_{cont} + L_{acc}$ is the sum of the energy gained through the contraction of the envelope $L_{cont}$ and by the accretion of planetesimals $L_{acc}$. The contraction luminosity, assumed to be constant throughout the envelope, is computed from the change of energy of the planet between the time $t$ and $t+dt$
\begin{equation}
   L_{cont} = - \frac{ E_{tot}(t+dt) - E_{tot}(t) - E_{gas,acc}}{dt}
\end{equation}
where $E_{tot}$ is the total planetary energy and $E_{gas,acc} = dt \dot{M}_{gas} u_{int}$ is the energy gained by the accretion of nebular gas with a specific internal energy $u_{int}$ at a rate $\dot{M}_{gas}$. The luminosity associated with the accretion of planetesimals, which are assumed to deposit their energy onto the core, can be written
\begin{equation}
   L_{acc} = G \frac{ \dot{M}_{core} M_{core}}{R_{core}}  
\end{equation}
where $\dot{M}_{core}$ is the mass accretion rate of the planetesimals which results in an increase in the core mass $M_{core}$ and radius $R_{core}$. It has to be noted that at $L_{cont}$ cannot be computed in a straightforward manner since in order to compute $E_{tot}(t+dt)$ the structure of the envelope at $t+dt$ needs to be known. This difficulty can be circumvented with the help of an iterative scheme \citep{fortieretal:2013}.

The internal structure equations are solved with four boundary conditions: 1) the radius of the core $R_{core}$, 2) the total radius of the planet $R_{M}$, 3) the surface temperature of the planet $T_{surf}$, and 4) the surface pressure $P_{surf}$. With these boundary conditions the structure equations provide a unique solution for a given planet mass. 

The core radius  can be calculated  (module 8 in Fig.~\ref{fig:schematic}) for a given core mass, composition (rocky or icy) and pressure at its surface (relevant for planets with a massive H/He envelope). AMB solve the internal structure equations for a differentiated core using a simple modified polytropic equation of state for the density $\rho$ as a function of pressure $P$ \citep{seageretal:2007}
\begin{equation}
   \rho(P) = \rho_{0} +  c P^{n}
\end{equation}
where $\rho_{0}$, $c$, and $n$ are material parameters. This EOS neglects the relatively small temperature dependency of $\rho$ for solids. Therefore it is sufficient to consider only the equations of mass conservation and hydrostatic equilibrium (Eqs. \ref{eq:masscons} and \ref{eq:hydrostaticequilibrium}) to calculate the core's internal structure and radius \citep[for details, see][]{mordasinietal:2012b}.  Regarding the composition, for rocky material, a silicate-iron ratio of 2:1 in mass is assumed as for the Earth, and the ice fraction is given self-consistently by the formation model as it is known whether the planet accretes rocky or icy planetesimals.

While the planet is embedded in the nebula, and the core is subcritical ($M_{\rm core}\lesssim10\mearth$) for gas accretion, the gaseous envelope of the proto-planet smoothly transitions into the background nebula. During this so-called attached phase, the total radius of the planet $R_M$ is given by \citep{lissaueretal2009}
\begin{equation}
   R_M = \frac{GM}{\frac{c^2_s}{k_1} + \frac{GM}{k_2 R_H}}
\end{equation}
where $c^2_s$ is the square of the sound speed in the mid-plane of the gaseous nebula at the location of the planet, $k_1$ and $k_2$ with values of $1$ and $1/4$ respectively. The temperature and the pressure at the surface of the planet  (module 6 in Fig.~\ref{fig:schematic}) are specified by a matching condition with the local properties of the disc 
\begin{equation}
   T_{surf} = \left( T^4_{disc} + \frac{3 \tau L}{16 \pi \sigma R_M^2} \right)^{1/4}
\end{equation}
\begin{equation}
   P_{surf} = P_{disc} (a)
\end{equation}
where $\tau= \kappa(T_{disc},\rho_{disc}) \rho_{disc} R_M$ \citep{papaloizouterquem:1999}, $L$ is the luminosity of the planet, and $T_{disc}, \rho_{disc}, P_{disc}$ are respectively the temperature, density, and pressure in the mid-plane of the disc at the planet's location. By solving these equations and requiring that pressure and temperature match the values of the nebula $R_M$ not only provides the internal structure but also the mass of the gaseous envelope. Hence, as long as the planetary envelope matches continuously the nebula at $R_M$, this procedure also determines the rate of gas accretion $\dot{M}_{gas}$ by comparison of the envelope mass at $t$ and $t+dt$.

As the core and/or envelope mass grows so that the planet becomes supercritical (its mass being larger than the critical mass), the gas accretion rate accelerates and eventually reaches a point when the disc can no longer sustain this rate. At this moment, the envelope detaches from the nebula. The planet's outer radius $R_{P}$ is no longer equal to the Roche limit but must be calculated. In fact, during this phase, the radius rapidly contracts from its original value down to $R_{P}\approx 2 \; \textrm{to} \; 5$ Jovian radii, depending upon the planet's entropy. In this detached phase, the planetary growth rate by gas accretion no longer depends on the planet's internal structure, but rather on the structure and evolution of the disc. For a 1+1D viscous disc, at a given time $t$, the radial mass flux at $r$ is given by:
\begin{equation}
   F(r) = 3 \pi \nu(r) \Sigma(r) + 6 \pi r \frac{\partial(\nu \Sigma)}{\partial r}
   \label{eq:massfluxindisk}
\end{equation}
Hence, the maximum mass delivery rate by the disc to the planet $\dot{M}_{gas,max}$ is given by the net mass flux entering and leaving the gas feeding zone of the planet $a_p \pm R_H$. This can be written as
\begin{IEEEeqnarray}{rCl}
   \dot{M}_{gas,max} & = & \max \left[ F(a_p + R_H),0 \right] \nonumber \\
                                  & & + \min \left[ F(a_p - R_H), 0 \right]
   \label{eq:mdotgasmax}
\end{IEEEeqnarray}
During a time $dt$, the maximum gas-mass that the proto-planet can accrete is given by $ \dot{M}_{gas,max} \times dt$. For simplicity, it is assumed that a fixed fraction \citep[0.75 to 0.9,][]{lubowdangelo:2006} of the disc's mass flux is accreted onto the planet.

Being detached from the disc, the continuity in pressure between the envelope and the disc is no longer a suitable boundary condition. In fact, both the ram pressure of the gas falling in from the boundary of the gas feeding zone to the planetary surface, where a standing shock is formed, and the photospheric pressure need to be accounted for. This provides the new pressure and temperature at the surface of the planet \citep[e.g.][]{bodenheimeretal:2000, papaloizounelson:2005} 
\begin{equation}
   P_{surf}= P_{disc} (a_p) + \frac{\dot{M}_{gas}}{4 \pi R_{P}^2} v_{ff} + \frac{2g}{3\kappa}
\end{equation}
where $g$ is the gravitational acceleration at $R_{P}$, and
\begin{equation}
   T^4_{surf}= (1-A) T^4_{neb} + T^4_{int}
\end{equation}
with $A$ the albedo and $v_{ff}$ the free-fall velocity from the limit of the boundary of the feeding zone to the surface given by
\begin{equation}
   v_{ff} = \sqrt{ \frac{2GM}{R_P} - \frac{2GM}{R_H}}
\end{equation}
and
\begin{equation}
   T^4_{int} = \frac{3 \tau L_{int}}{8 \pi \sigma R^2_{P}}
\end{equation}
where $\tau =\max \left[\kappa(T_{disc},\rho_{disc}) \rho_{disc} R_P, 2/3\right]$.

These boundary conditions basically envision a miniature version of spherical accretion onto a stellar core as in \citet{stahleretal:1980}. The actual geometry, and therefore also the boundary conditions are more complex and require in principle 3D radiation-hydrodynamic simulations \citep[e.g.][]{klahrkley:2006}. For the calculation of $L_{int}$, assumptions have to be made regarding the structure of the accretion shock: if the shock is radiatively (in)efficient, the potential energy of the gas liberated at the shock is (is not) radiated away, so that material gets incorporated into the planet at low (high) entropy, resulting in a low (high) luminosity \citep[e.g.][]{marleyetal:2007,spiegelburrows:2012,mordasinietal:2012}.  Typically, the limiting case of completely cold/hot accretion are considered. Together with D-burning in more massive objects \citep[][]{spiegeletal:2011,mollieremordasini:2012,bodenheimeretal:2013}, this yields the post-formation luminosity (and radius) of giant planets (hot/cold start) which is crucial for the interpretation of directly imaged planets.

The final evolutionary (or isolated) phase occurs after the proto-planetary nebula has dissipated so that the planet cools and contracts, conserving the total mass (neglecting further accretion or mass loss, e.g., through atmospheric escape for close-in planets). The simplest possibility to model this phase is with a gray atmosphere so that
\begin{equation}
   P_{surf}= \frac{2g}{3\kappa}
\end{equation}
where the opacity $\kappa$ is now given by the grain-free opacities of \citet{freedmanetal:2008}, and
\begin{equation}
   T^4_{surf}= (1-A) T^4_{eq} + T^4_{int}
\end{equation}
with $T_{eq}$ the equilibrium temperature due to stellar irradiation (Eq.~\ref{eq:temp_dist}). As noted by \citet{bodenheimeretal:2000} these simple models lead to luminosities and radii as a function of time that agree relatively well with full non-gray models \citep[e.g.][]{burrowsetal:1997,baraffeetal:2003}. This enables us to compare the radii (and luminosities) calculated by synthetic populations with results of transit (and direct imaging) surveys.

While the approach used by IL is computationally extremely rapid and therefore allows the tests of many models, it includes uncertainties in the mass and structure of planets, especially Earth-like or super-Earth planets with small mass atmospheres. On the other hand, it should also be pointed out that full computations by AMB are only accurate as long as the ingredients used are well justified, which are not always the case. For example, the EOS, the opacity, the structure of the accretion shock} \citep[e.g.][]{marleyetal:2007} or the models of convection \citep[e.g.][]{baraffeetal:2012, lecontechabrier:2012} are uncertain.

\subsection{Interactions between growing planets}
\label{subsec:interactionsgp}
%------------------------------------------------------------------

\noindent Interactions between planets growing within a given proto-planetary disc are key ingredients in a model of the formation of planetary systems (module 10 in Fig.~\ref{fig:schematic}). Because eccentricity damping due to dynamical friction with the gaseous disc \citep[e.g.][]{tanakaward:2004} is very strong, proto-planets formed through oligarchic growth are usually isolated from each other and in nearly circular orbits. Once the disc of gas is sufficiently depleted, secular perturbations pump up eccentricities \citep{chambersetal:1996} leading to orbit crossing and hence to collisions. Because a single collision can double the mass of the bodies involved, planetary masses can eventually increase by orders of magnitude. On the other hand, scattering among gas giant planets often results in ejection of one or two planets, leaving other planets in highly eccentric isolated orbits \citep[e.g.][]{marzariweidenschilling:2000, nagasawaetal:2007}. The perturbations from gas giants on such eccentric orbits can alter orbital configurations of a whole planetary system. Even rocky planets in the inner regions that are far from the interacting gas giants in outer regions can be severely affected \citep[e.g.][]{matsumuraetal:2013}.  

In early population synthesis models, interactions between planets have been neglected, because they are chaotic and highly non-linear and hence difficult to model. However, recent models \citep{idalin:2010, idaetal:2013, alibertetal:2013} have now incorporated planet-planet gravitational interactions.

 Because explicit N-body simulations are computationally very expensive, IL developed semi-analytical Monte-Carlo models to compute planet-planet collisions and scatterings. Their approach is based on detailed planetesimal dynamics studies revealed by prior detailed N-body simulations and statistical formulations  \citep[e.g.][]{idanakazawa:1989, ida:1990, idamakino:1993, palmeretal:1993, aarsethetal:1993, kokuboida:1998, kokuboida:2002, ohtsukietal:2002}. As a result, their models reproduce quantitatively statistical distributions of N-body simulation outcomes. Because the prescriptions of planet-planet collisions and scatterings are rather complicated multi-step schemes, we omit the descriptions of their prescriptions and refer for details to \citet{idalin:2010, idaetal:2013}.

In contrast, AMB's combine direct N-body simulations with population synthesis calculations, in order to accurately take into account the effects of planet-planet interactions. These authors use a  standard integration scheme (e.g., Bulirsch-Stoer). The equation of motions for the proto-planets in a heliocentric reference frame are written as
\begin{IEEEeqnarray}{rCl}
   \ddot{ \vect{r_i}} & = & - G \left(M_* + m_i \right) \unitvq{r_i} \nonumber \\
                              && - G \sum_{j=1,j\neq i}^{n} {m_j \left\{ \frac{\vect{r_i}-\vect{r_j}}{\left| \vect{r_i}-\vect{r_j} \right|^3} + \unitvq{r_j} \right\} }
\end{IEEEeqnarray}
with $ i = 1,2,3 \dots N$, $m_i$ the mass of the planets and $M_*$ the mass of the central star. The integration uses an adaptive times-step which ensures reaching a desired precision. We note that since the integration is carried out over the proto-planetary disc lifetime ($\le$ 10 Myr) and that dissipative forces exist, a simplectic integrator is not necessary. While this approach is relatively straight forward it has the disadvantage of being relatively expensive in integration time for a large number of proto-planets and/or for small time steps. It has, however, the great merit to capture all possible effects (e.g., resonances, collisions). 

Collisions between the planets are detected by checking if two planets come closer to each other than $d_{col}=R_1 + R_2$ where $R_1$ and $R_2$ are the radii of the two bodies. This is achieved by searching among all possible pairs of proto-planets those which, during the time $t$ and $t+ \Delta t$, will come closer than $d_{col}$ to each other; $\Delta t$ being the time step of the N-body integrator. In practice, this is done by extrapolating the positions of all proto-planets using Taylor expansions, and search for a time $\tau$ such that for $t \le \tau \le t +\Delta t$ the distance between any possible pair of proto-planets is less than $d_{col}$. This approach has been proposed by \citet{richardsonetal:2000} who used a first order Taylor expansion and has more recently been improved by  \citep{alibertetal:2013} by using second order extrapolations. 

\subsection{Planet-disc interactions: migration}
\label{subsec:migration}
%------------------------------------------------------------------

\noindent  Planet-disc interactions lead to planet migration and to the damping of eccentricity and inclination. The migration results from planet-disc angular momentum exchange and the determination of the migration rate requires the computation of the 2D or 3D structure (including the thermodynamics) of the proto-planetary disc. The computer time required for carrying out such detailed multi-dimensional simulations of these processes over a timescale covering planet formation is prohibitively high. Hence, in the population synthesis approach, planetary migration is computed using fits to migration rates resulting from hydrodynamical calculations (see Chapter by Baruteau et al. and references therein).

Planetary migration occurs in different regimes depending upon the mass of the planet. For low mass planets, \textit{i.e.\ }Êplanets not massive enough to open a gap in the proto-planetary disc, migration occurs due to the imbalance between the Lindblad and corotation torques exerted on the planet by the inner and outer regions of the disc. This regime is called "type I migration" and the corresponding migration rate has been derived by linear and numerical calculations \citep[e.g.][]{ward:1997, tanakaetal:2002, paardekooperpapaloizou:2009, paardekooperetal:2011}. For higher mass planets, \textit{i.e.\ } planets massive enough to open a gap in the proto-planetary disc, the planet is confined in the gap by Lindblad torques and thus follows the global disc accretion. This regime is called "type II migration."  Type II migration is itself sub-divided in two modes: disc-dominated type II migration, in the case the local disc mass exceeds the planetary mass, and planet-dominated type II migration in the opposite case  \citep[see also][]{linpapaloizou:1986, idalin:2004a, mordasinietal:2009a}. In the former case, the migration rate is simply given by the local viscous evolution of the proto-planetary disc, while the migration is decelerated by the inertia of the planet in the latter case.

Initial population synthesis models by both IL and AMB made use of  the conventional formula of type I migration derived for locally isothermal discs \citep{tanakaetal:2002}. In order to investigate how sensitive the results are on the magnitude of this migration, a scaling factor $C_1$ was introduced: 
\begin{IEEEeqnarray}{rCl}
   \tau_{\rm mig1} & = & \frac{a}{\dot{a}}\nonumber \\
                             & = & \frac{1}{C_1} \frac{1}{3.81} \left(\frac{c_s}{a \Omega_{\rm K}}\right)^{2}  \frac{M_*}{M_{\rm planet}} \frac{M_*}{a^2 \Sigma_g}
                                        \Omega_{\rm K}^{-1} \nonumber \\
                             & \simeq & 1.5 \times 10^5 \frac{1}{C_1 f_g}  \left(\frac{M_{\rm c}}{M_{\oplus}} \right)^{-1}   \left(\frac{a}{1{\rm AU}}\right) \nonumber \\
                             & &  \times \left(\frac{M_*}{M_{\odot}}\right)^{3/2} \;{\rm yrs}. 
 \label{eq:tau_mig1} 
\end{IEEEeqnarray} 
The expression of \citet{tanakaetal:2002}\ corresponds to $C_1 = 1$, while $C_1 < 1$ implies slower migration rates. IL assume type I migration  ceases inside the inner boundary of the disc. 

Since the publication of \citet{tanakaetal:2002}, radiative effects on the type I migration rate have been investigated \citep[e.g.][]{paardekoopermellema:2006,massetcasoli2010, paardekooperetal:2011}. It was shown that the migration velocity as well as its direction depend sensitively upon the detailed dynamical and thermal structure of discs, leading to a number of sub-regimes of type I migration (locally isothermal, adiabatic, (un-)saturated). Recently, a new semi-analytic description of type I migration, which reproduces the results of \citet{paardekooperetal:2011}, has been derived \citep[][]{mordasinietal:2011,kretkelin:2012}. It includes the effect of co-rotation torques that can lead to outward migration in non-isothermal discs. This new formalism has been implemented in recent simulations by AMB who have shown that the scaling factor determining the migration speed introduced in earlier models (Eq.~\ref{eq:tau_mig1}) becomes much less important \citep{alibertetal:2013}. 

The transition mass between type I and type II migration is $M_{\rm g,vis}$ (Eq.~\ref{eq:m_gas_vis}) in IL's prescription and $M_{\rm g,th}$ (Eq.~\ref{eq:m_gas_th}) (more exactly, the condition derived by \citealt{cridaetal:2006}) in AMB's prescription. The comparison between gap opening criteria is treated in more details in section \ref{subsec:gap}.

Initially, population synthesis models have assumed an isothermal migration rate reduced by  $C_1 \sim 10^{-2}-10^{-1}$ in IL's simulations and $C_1 \sim 10^{-3}-10^{-2}$ in AMB's simulations. The values of $C_1$ less than unity were needed to prevent cores of growing giant planets to fall into the host star. These findings by population syntheses were an important motivation to develop physically more realistic  non-isothermal migration models. This is one out of several examples in which population synthesis can be used to test in a statistical sense detailed modelling of individual processes. Population synthesis models did not provide a better understanding of the migration itself but pointed out that the current prescription did not result in planet populations with the observed characteristics. It is worth pointing out that with this new formalism for type I migration, which has been implemented in recent simulations by AMB,  an arbitrary scaling factor (Eq.~\ref{eq:tau_mig1}) slowing down migration is no longer an absolute necessity  \citep{alibertetal:2013}. While this represents a definitive progress, difficulties remain. They are linked to the sensitivity of the migration rate to the saturation of the corotation torque, to a partial gap formed by relatively large migrating planet, and to orbital eccentricity. A further difficulty is due to the fact that the onset of efficient gas accretion onto the core, and the saturation of the corotation torque occur at a similar mass (of order 10 $\mearth$), so that a self-consistent coupled approach of the two processes is necessary. 

For type II migration, as long as the mass of the planet remains smaller than the local disc mass (of the order of $\pi \Sigma r^2$), the migration timescale ($\tau_{\rm mig2}(a) =a/|v_r|$) is given by the local viscous diffusion time, $t_{\rm vis}(a) \sim (2/3)(a^2/\nu)$. For a steady accretion disc with $F \sim 3 \pi \Sigma \nu$ and $\Sigma \propto 1/a$, $M_{\rm disc}(a) = \int^a 2\pi a \Sigma da = 2\pi \Sigma a^2$. Then, $\tau_{\rm mig2}(a) \sim t_{\rm vis}(a) \sim M_{\rm disc}(a)/F$ \citep{hasegawaida:2013}. For a planet more massive than the inner disc mass $M_{\rm disc}(a)$, the viscous torque from the outer disc pushes the planet (mass $M_{\rm planet}$) rather than the inner disc. Then, replacing $M_{\rm disc}$ with $M_{\rm planet}$, $\tau_{\rm mig}(a) \sim M_{\rm planet}/F$ \citep{hasegawaida:2013}. In summary, the type II migration rate ($v_r \sim a/\tau_{\rm mig2}$) is roughly given by
\begin{equation}
  \frac{d a}{d t} = -\frac{3 \nu}{2 a} \times \min \left[1,\frac{M_{\rm disc}(a)}{M_{\rm planet}}\right].
   \label{eq:type2_simple}
\end{equation}
Migrations for $M_{\rm planet} < M_{\rm disc}(a) $ and $M_{\rm planet} > M_{\rm disc}(a)$  are called "disc-dominated" and "planet-dominated" type II migrations, respectively.

IL and AMB use more detailed prescriptions for planet-dominated regime. IL adopt \citep{idalin:2008a}
\begin{IEEEeqnarray}{rCl}
   \frac{d a }{d t} & = &  -\frac{3 \nu}{2 a} \nonumber \\
                          & & \times \min \left[1,2C_2\left( \frac{30{\rm AU} }{a}\right)^{1/2}\frac{M_{\rm disc}(a)}{ M_{\rm planet}}\right].
\end{IEEEeqnarray}
Because they use $C_2 = 0.1$ and usually $a \sim O(1)$, the extra factor $2C_2(30{\rm AU}/ a)^{1/2}$ is $\sim 1$ and IL's prescription is the same as the simple formula given by Eq.~(\ref{eq:type2_simple}).

AMB use
\begin{equation}
   \frac{d a}{dt} = -\frac{3 \nu}{2 a} \times \min \left[1,\frac{2 \Sigma a^2}{M_{\rm planet}}\right].
\end{equation}
In the region of $\Sigma \propto 1/a$, this formula becomes
\begin{equation}
  \frac{d a}{d t} = -\frac{3 \nu}{2 a} \times \min \left[1,\frac{1}{\pi} \frac{M_{\rm disc}(a)}{M_{\rm planet}}\right].
\end{equation}
Thus, AMB's type II migration rate is slower than IL's by a factor of $\pi$ in the planet-dominated regime, while the rates are identical in the disc-dominated regime. More detailed hydrodynamical simulations are required to find which migration rate is more appropriate in the planet-dominated regime. However, as discussed below, different treatments of gas accretion onto planets after gap formation affect more significantly the efficiency of type II migration in population synthesis simulations.

\subsection{Planet-disc interactions: gap formation}
\label{subsec:gap}
%--------------------------------------------------------------------

\noindent  Gap formation in a gas disc by the perturbations exerted by a planet has two important consequences: 1) it reduces or even terminates the gas flow onto the planet and 2) the planet switches its migration from type I to type II. While gravitational torques exerted by the planet on the disc work towards the opening of a gap, two physical processes tend to prevent this opening: viscous diffusion and pressure gradients. The planet's tidal torque exceeds the disc's intrinsic viscous stress  at the mass \citep{linpapaloizou:1986},
\begin{IEEEeqnarray} {rCl}
   M_{\rm planet}  > M_{\rm g,vis} & \simeq  & \frac{40 M_\star}{Re} \nonumber \\
                                                      & \simeq  &  40 \alpha \left(\frac{H_{\rm disc}}{a}\right)^2 M_\star \nonumber \\
                                                      & \simeq & 30 \left(\frac{\alpha}{10^{-3}}\right) \left(\frac{a}{1{\rm AU}}\right)^{1/2} \nonumber \\
                              && \times \left(\frac{L_\ast}{L_{\odot}}\right)^{1/4} M_{\oplus},
   \label{eq:m_gas_vis} 
\end{IEEEeqnarray} 
where $H_{\rm disc}$ is the disc scale-height at the location of the planet, and $Re = a^2 \Omega / \nu$ is the Reynolds number at the location of the planet ($a$). This is called the "viscous condition" for gap formation. If the gap half width is less than $\sim H_{\rm disc}$, the pressure gradient inhibits gap formation. The gap half width may be $\sim R_{\rm H}$ where $R_{\rm H}$ is the planet's Hill radius. Then the critical mass is given by $R_{\rm H} \ga  \beta H_{\rm disc}$, where $\beta \sim O(1)$, that is,
\begin{IEEEeqnarray}{rCl}
   M_{\rm planet} > \beta^3 M_{\rm g,th} & \simeq &  3 \left(\frac{\beta H_{\rm disc}}{r}\right)^3 M_\star \nonumber \\ 
                                                                & \simeq & 120 \beta^3 \left(\frac{a}{1{\rm AU}}\right)^{3/4} \left(\frac{L_\ast}{L_\odot}\right)^{3/8} \nonumber \\
                                                                 & & \times \left(\frac{M_\ast}{M_\odot}\right)^{-1/2} M_\oplus,
   \label{eq:m_gas_th}
\end{IEEEeqnarray}
 where $M_{\rm g,th}$ is defined by $3(H_{rm disc}/r)^3M_{\odot}$
(note that the definition is different from that in IL's papers by a factor of $\beta^3$).
This is called the "thermal condition."

\citet{LinPapaloizou:1993,cridaetal:2006} found through numerical calculations that the gap opening condition is 
\begin{equation}
   \frac{3}{4} \frac{ H_{\rm disc} }{R_{\rm H}} + \frac{50 M_\star }{ M_{\rm planet} Re} < 1,
\label{eq:Crida}
\end{equation}
which is equivalent to a combination of viscous condition and thermal condition (with $\beta=3/4$).

IL adopt the following prescription:
\begin{enumerate}
   \item For $M_{\rm planet} > M_{\rm g,vis}$ $(\beta=2)$, a gap is formed and type I migration is switched to type II migration. Here, a gap is assumed to be partial (low density region along the planet's orbit), so that the gas disc still crosses the gap.
  \item For $M_{\rm planet} > 8 M_{\rm g,th}$, gas accretion onto the planet is completely terminated, because hydrodynamical simulations show that gas-flow across the gap rapidly decays as $M_p$ increases beyond Jupiter mass \citep[e.g.][]{dangeloetal:2002, lubowdangelo:2006} and \citet{dobbs-dixonetal:2007} suggests that gas accretion onto the planet is terminated if $R_{\rm H}$ well exceeds $H_{\rm disc}$.
  \item As a result, the accretion rates onto the planet is given by ($F$ is the mass flux in the disc, Eq.~\ref{eq:massfluxindisk}).
\begin{equation}
   \frac{dM_{\rm planet}}{dt} = f_{\rm gap} F,
\end{equation}
where $f_{\rm gap}$ is a reduction factor due to gap opening,
   \begin{equation}
   f_{\rm gap} = 
   \left\{
   \begin{array}{ll}
   1 & [M_ p <  M_{\rm g,vis}] \\
    0 & [M_ p >  8M_{\rm g,th}],
   \end{array}
   \right.
\end{equation}
and for $M_{\rm g,vis} < M_ p <  8M_{\rm g,th}$
\begin{equation}
   f_{\rm gap} = \frac{\D \log M_{\rm planet} - \log M_{\rm g,vis}}{\D \log 8M_{\rm g,th} - \log M_{\rm g,vis}}
\end{equation}
 This formula is constructed to avoid any abrupt truncation.  
 \item The gas accretion also decays according to global disc depletion. This effect is automatically included through $F$ that is proportional to $\Sigma$.
\end{enumerate}

AMB adopt a different prescription.
\begin{enumerate}
\item When the condition (\ref{eq:Crida}) is satisfied, a gap is formed and the migration mode switches from type I to type II. Note that for $\beta = 3/4$, the thermal condition $M_{\rm g,th} > (3/4)^3M_{\rm g,th}$ is comparable to the viscous condition $M_{\rm planet} > M_{\rm g,vis}$. 
\item After gap opening and provided the planet is in the detached phase, the gas accretion onto the planet is remains given by $F$ times a fixed factor $<1$ (see section \ref{subsec:gasaccretion})  without any further reduction. The planet keeps growing until disc gas becomes globally depleted, which manifests itself by a gradual decrease of $F$ to zero. This limiting assumption of no reduction due to gap formation is motivated by the results of  isothermal hydrodynamic simulations of \citet{kleydirksen:2006}. They found  that for planetary masses above a certain minimum mass ($3-5$ Jovian masses, depending upon viscosity), the disc makes a transition from a circular state into an eccentric state. In this state, the mass accretion rate onto the planet is greatly enhanced relative to the case of a circular, clean gap because the edge of the gap periodically approaches the proto-planet that can even become (re-)engulfed in the disc gas for large eccentricities.
\end{enumerate}

Because AMB assume that the planet continues accreting gas at the unimpeded disc accretion rate (Eq.~\ref{eq:massfluxindisk}) even well after gap opening, and consider relatively efficient photo-evaporation (which is not considered by IL), the migration soon enters the planet-dominated regime and becomes slower and slower as the planet keeps growing. With AMB's prescription for planet-dominated regime,  it can be shown that $d\log M_p/d\log a = (a/M_p)(dM_p/dt)/(da/dt) \sim  -\pi$.  This is clearly in contrast to IL's approach described above. As a consequence, even though the type II formalism is very similar, the predicted distributions of semi-major axis of gas giants between the two approaches are different. This is essentially due to the increased role of planet inertia in the AMB formalism which slows down the migration of massive planets. This leads IL to predict a much larger frequency of hot Jupiters (\ref{subsec:compa}).

\bigskip
\section{INITIAL CONDITIONS}
\label{sec:ic}
\bigskip

The single most important ingredient of the population synthesis method is a global planet formation model that ``translates'' properties of a proto-planetary disc (which are the initial conditions for planet formation) into properties of the emerging planetary system. This formation model has been described in the previous section. The other most important ingredients are the probabilities of occurrences (distributions) of these initial conditions.

The initial conditions of population synthesis calculations are of two different types. The first one is related to the properties of the proto-planetary disc while the second is related to the properties of the proto-planets themselves. As far as the foist type is concerned, the basic assumption of planetary population synthesis is that the (observed) diversity of planetary systems is a consequence of the (observed) diversity of the properties of proto-planetary discs. This assumption is verified \textit{a posteriori} by the large diversity of planets resulting from this assumption. The initial conditions characterising a disc are therefore treated as Monte Carlo variables that can be drawn from probability distributions.

Ideally, the probability distributions of the properties of proto-planetary discs should be taken directly from observations (see Chapters by Dutrey et al. and  Testi et al.). Unfortunately, this is not a straightforward task, as present day observations do not constrain the innermost parts of discs (where planets actually form) very well, and the number of well characterised discs is presently small. Both IL and AMB consider three fundamental disc properties as Monte Carlo variables: 
\begin{enumerate}
  \item The (initial) surface density of gas in the proto-planetary disc. In the IL models (Eq.~\ref{eq:sigma_gas}),  it is given by the scaling factor $f_{g,0}$ (a scaling factor for $\Sigma_g(r=\rm{10 AU},t=0)$). According to the distributions of total disc masses inferred by radio observations of T Tauri discs, IL assume a Gaussian distribution of $\log_{10} f_{g,0}$ with a mean and standard deviation of 0 and 1 respectively. In the AMB models, the different disc masses are represented by different $\Sigma(r=\rm{5.2 AU},t=0)$ in Eq.~(\ref{disk_surf_evol}).  For the distribution of disc masses, AMB fit the disc mass distribution observed by \citet{andrewsetal:2010} with a Gaussian distribution with a mean and standard deviation in $\log_{10}(M_{\rm disc}/M_{\odot})$ of -1.66 and 0.56, respectively, or alternatively directly boot-strap from the observed distribution \citep{fortieretal:2013}. Note that the mean value, $\log_{10}(M_{\rm disc}/M_{\odot})= -1.66$, is comparable to the disc mass corresponding to $\log_{10} f_{g,0} \sim 0$ with the disc size $\sim 100$AU.

   \item The lifetime of the proto-planetary disc. In the IL models (Eq.~\ref{eq:gas_exp_decay}), it is represented by $\tau_{\rm dep}$. According to IR and radio observations, IL assume a Gaussian distribution of $\log_{10} \tau_{\rm dep}$ with a mean and standard deviation 6.5 and 0.5. In the AMB models, the distribution of disc lifetimes is obtained by specifying a distribution of external photo-evaporation rates (Eq.~\ref{eq:BAMextphotoevap}). This distribution is adjusted  in a way \citep[for details, see][]{mordasinietal:2009a} that the distribution of the resulting lifetimes of the synthetic discs agrees with the observed distribution as derived from the fraction of stars with an IR excess \citep{haischetal:2001} (see also Chapter by Dutrey et al.)
   
     \item The surface density of solids. In the IL models (Eq.~\ref{eq:sigma_dust}), it is represented by the scaling factor $f_{d}$. The initial value, $f_{d,0}$, is given by $f_{g,0}10^{\rm [Fe/H]}$, where [Fe/H] is the stellar metallicity and the same dust-to-gas ratio is assumed between the stellar surface and the disc. Consequently, the initial dust-to-gas ratio $f_{D/G}$ is given by $f_{D/G,0}(f_{d,0}/f_{g,0})$, where $f_{D/G,0}$ is the ratio associated with the solar composition (IL adopt $f_{D/G,0} = 1/240$).  In the AMB models, $f_{D/G}=f_{D/G,0}10^{\rm [Fe/H]}$ is first specified and then the surface density of solids is given by Eq.~(\ref{eq:sigmasolidBAM}). Thus, the IL and AMB models are equivalent as far as the description of the surface density of solids is concerned. For the probability distribution, IL use a Gaussian distribution of [Fe/H] with a mean and standard deviation of 0 and 0.2 dex, respectively, while AMB use that of -0.02 and 0.22 dex, corresponding to the CORALIE planet search sample \citep{udryetal:2000}.
 
\end{enumerate}

Besides the total disc mass, it is also necessary to specify the radial profile of the gas (and solid) surface density. One approach is to assume some theoretically inspired surface density profile, mass and composition. First models of population synthesis have indeed assumed disc profiles similar to the minimum solar nebula, with a  surface density slope of typically $-3/2$.

Recently, as the number of well characterised discs has grown, new models started to consider disc profiles that come directly from fits of observations \citep[e.g.][]{andrewsetal:2010}. Typical disc profiles are given by:
\begin{IEEEeqnarray}{rCl}\label{disk_andrews}
\Sigma (r)  & = & (2 - \gamma) \frac{ M_{\rm disc} }{ 2 \pi a_{\rm core} ^{2-\gamma} r_0 ^\gamma } \left( \frac{r }{ r_0} \right)^{-\gamma} \nonumber \\
            & & \times \exp \left[ - \left( \frac{r }{ a_{\rm core}} \right) ^{2-\gamma} \right]
,
\end{IEEEeqnarray}
where $r_0$ is equal to 5.2 AU, and $M_{\rm disc}$, $ a_{\rm core}$, $\gamma$ are derived from observations. The observations \citep[e.g.][]{andrewsetal:2010} suggest $\gamma\sim 1$, which is consistent with the self-similar solution with constant $\alpha$. Accordingly, IL adopt $\Sigma(r) \propto 1/r$ in recent models, as already mentioned.  AMB, initially adopted  $\Sigma(r) \propto r^{-1/5}$. In more recent models, the gas surface density is taken to follow the distribution of Eq. \ref{disk_andrews}, the disc parameters being directly the ones derived in \cite{andrewsetal:2010}.

To start a population synthesis calculation it is necessary to distribute planetary seeds within the disc. The growth of these seeds is then followed in time. The initial location and mass of these seeds are not constrained from observations and are derived from theoretical arguments. Seeds are often assumed to have an initial location distribution that is uniform in log, following N-body calculations of the early stage of planetary growth \citep[e.g.][]{kokuboida:1998,kokuboida:2002}. IL use a somewhat different approach. The masses of proto-planets formed by oligarchic growth are predicted by $\Sigma_{\rm s}$ distribution as "isolation" masses in inner regions or final masses predicted by simple formula in outer regions (where planetesimal accretion is so slow that the proto-planets' mass does not reach their isolation mass), so that seeds are set up with orbital separations comparable to feeding zone width of the proto-planets.

One should note that recent models of planetesimals and proto-planet formation (see Chapters by Johansen et al., Raymond et al., and Helled et al.) predict that proto-planets might form under precise circumstances (e.g., close to the ice line or near a pressure maximum). Under these circumstances, the distribution of initial locations of proto-planets would be far from a uniform in log-scale but rather 
concentrated at specific locations in the disc. A better understanding of these issues is necessary for future progress in population synthesis models. 

The initial mass of the seeds is similarly observationally undetermined. Ideally, the result of population synthesis models should be independent of the assumed initial mass of seeds, provided the initial proto-planetary disc model is consistent with the time required to grow these seeds. More massive seeds taking longer to grow should be implanted in already more evolved discs. This obviously raises the question of how to define time zero. In all cases, the initial mass of seeds should be small enough so that any processes such as migration or gas accretion remains negligible. In practice, it is often a good choice to assume that the initial mass of seeds corresponds to the one obtained at the end of the local runaway growth phase, as this latter is believed to be very rapid.

\bigskip
\section{OBSERVATIONAL BIASES}
\bigskip

One of the key objectives of population synthesis models is to compare models with observations, if possible even in a quantitative way. From this comparison, constraints on some of the key processes acting during planet formation should be gained. However, for this comparison to be meaningful, it is essential that observational selection biases associated with the exoplanet detection methods are well understood. Large, homogeneous surveys with a well characterised detection bias like the HARPS or Eta-Earth surveys for the radial velocity technique \citep{mayoretal:2011, howardetal:2010}, or the Kepler mission for the transit technique \citep{boruckietal:2011} are, in this respect, of particular interest. Since the majority of exoplanets have been discovered by these two techniques, their selection biases are also best known. For radial velocities, the simplest approach is to consider a detection criterion based on the induced radial velocity amplitude and a maximal period. A more sophisticated approach uses a tabulated detection probability that is a function of the planet's mass and orbital period, and takes the instrumental characteristics as well as the actual measurement schedule into account \citep{mordasinietal:2009b}. In addition, recent  ``controlled experiment'' microlensing surveys  \citep{gouldetal:2010} and forthcoming large direct imaging (with GPI at the Gemini Observatory, and SPHERE at the Very Large Telescope) and astrometric surveys (\textit{e.g.}, GAIA) are equally important, as different techniques typically probe different sub-population of planets, yielding complementary constraints for the modelling. It would be particular informative if the systematic surveys can provide well determined upper limits on both the presence and absence of planets in domains of parameter space.  Such data can be used to verify or falsify predictions on different planetary characteristics made with population synthesis models.

\bigskip
\section{{\bf HIGHLIGHTS OF POPULATION SYNTHESIS MODELS}}
\label{sec:results}
\bigskip

\subsection{Individual systems}
%------------------------------------------

\noindent  Population synthesis predictions are made by the superposition of individual systems with different initial conditions (section \ref{subsec:compa}). Before we start statistical discussions on the distributions predicted by population synthesis in comparison with the observed ones, we first show examples of the evolution of individual systems.

\begin{figure*}[htp]
\epsscale{1.7}       
 \plotone{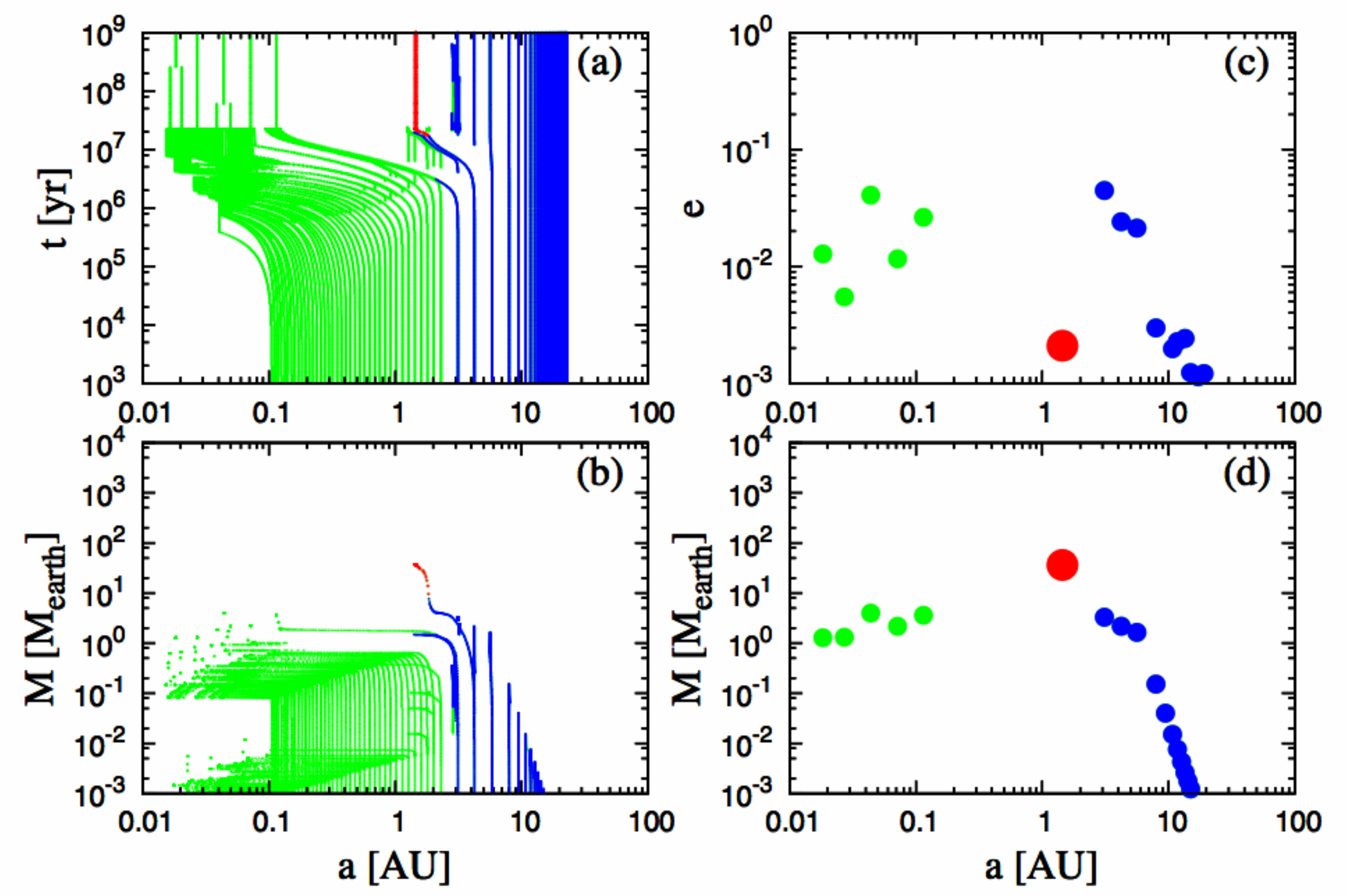} 
\caption{\small  
Growth and migration of planets in a system with $C_1 = 0.1$ and $f_{d,0}=2.0$. The green, blue and red lines represent planets with their main component is rock, ice and gas, respectively. 
Panels (a) and (b) show time and mass evolution of planets with different initial semi-major axes. Panels (c) and (d) are orbital eccentricity and (d) planetary mass as a function of semi-major axis in the final state.
[from \citet{idaetal:2013}]
}
  \label{fig:obt}
\end{figure*}

Fig.~\ref{fig:obt} shows an example of evolution of a system using the IL approach with $C_1=0.1$, $f_{g} = f_{d} = 2.0$ ([Fe/H]$=0$), $\tau_{\rm KH1}=10^9$ years,  $\tau_{\rm dep} = 3 \times 10^6$ years and orbiting a solar-mass star ($M_* = 1 M_{\odot}$). Panels a and b show time and mass evolution.  The green, blue and red lines represent rocky, icy, and gas giant planets with their main component being rock, ice and gas, respectively.  The bulk composition of the planets can change over time through gas or planetesimal accretion or planet-planet collisions. At 0.1--1AU small proto-planets grow {\it in situ} until their masses reach $\sim 0.1M_{\oplus}$ --$1M_{\oplus}$ and then they undergo type I migration and accumulate near the inner boundary of the disc which is set at 0.04 AU. Many resonant proto-planets actually accumulate at the vicinity of this boundary (Panel a) and are preserved until the gas disc decays enough to allow orbit crossing and merging starts. Just outside the ice line at $a \sim 3$AU, a core reaches $M_p \sim 5 M_{\oplus}$ and starts runaway gas accretion without any significant type I migration.  After it has evolved into a gas giant with a surrounding gap, it undergoes type II migration.  The emerging gas giant scatters and ejects nearby proto-planets. Finally, a system with closely-packed close-in super-Earths, a gas giant at an intermediate distance, and outer icy planets in nearly circular orbits is formed (Panel c and d). Since the super-Earths are formed by scattering and merging, they have been kicked out from resonances. Such closely-packed, non-resonant, close-in super-Earths are found to be common by Kepler observations.

Note that in our Solar system, no planet exists inside of Mercury's orbit at $a = 0.39$AU, while RV and Kepler observations suggest that more than 50\% of solar-type stars have close-in planets. Hence, a planetary system like our own solar system may only form if either we loose these inner most planets to the sun (no inner cavity exist) or outward type I migration took place. Furthermore, for giant planets such as Jupiter and Saturn to remain at large distances, they must have formed late, at a time when the disc was severely depleted in order to avoid extensive type II migration. Since a significant time lag may exist between the formation of two giant planets, it is not easy for the current model to explain the presence of two gas giants in the outer regions. A mechanism to form two gas giants almost simultaneously might be required such as in the induced formation model by \citet{kobayashietal:2012}. (Such an effect has not yet been incorporated into population synthesis simulations.)

\subsection{Comparisons with observations}
\label{subsec:compa}
%-----------------------------------------------------------

\noindent  In this section we present some aspects of the statistical comparisons between synthetic populations and observations. To this effect, a population of planets is built by running the planet formation model using a large number of different initial conditions. As explained above, the initial conditions are drawn at random following a Monte Carlo procedure in which observations and theoretical arguments are used to determine the probability of occurrence of a given initial condition. 

One of the key result of population synthesis models is the computation of the mass \textit{versus} semi-major axis diagram of planets. Such a diagram might be of similar importance for planetary physics than the Hertzsprung-Russell diagram for stellar astrophysics. First models by \citet{idalin:2004a}, and later by \citet{mordasinietal:2009b} provided a consistent global picture, with however some interesting differences. Indeed, both models concluded that type I migration had to be highly reduced compared to the theoretical estimates available at the time \citep[e.g.][]{tanakaetal:2002}. Moreover, it appeared naturally from the very concept of the core-accretion model that there should be a lack of planets of masses between Neptune and Saturn. This results simply from the fact that the accretion of gas is low for sub-critical planets (with a core mass less than $\sim 10 \mearth$), whereas it is quite rapid for super-critical planets \citep{pollacketal:1996}, up to the point when additional processes hinder gas accretion (for masses larger than $\sim 100 \mearth$). Since it is unlikely that the proto-planetary disc, and therefore the gas supply disappears exactly during the short timescale of gas runaway accretion, less planets with intermediate masses are expected. \citet{idalin:2004a} called this potential deficit of intermediate mass planets the "planetary desert."

How desert the "planetary desert" actually is depends upon the details of the computation of the gas accretion rate \citep[for a dedicated discussion, see][]{mordasinimayor:2011}. In particular, the AMB models and \citet{idaetal:2013} limit the gas accretion rate onto the planet by the rate at which the disc can actually provide gas to the planet. This modification reduces the gas accretion rate, and the "planetary desert" becomes less pronounced than in the earlier models by \citet{idalin:2004a}. This difference shows that the comparison of the actual and synthetic $a-M$ diagram helps to better understand the mechanism of gas accretion. 

\begin{figure}[htp]
 \epsscale{1.0}
 \plotone{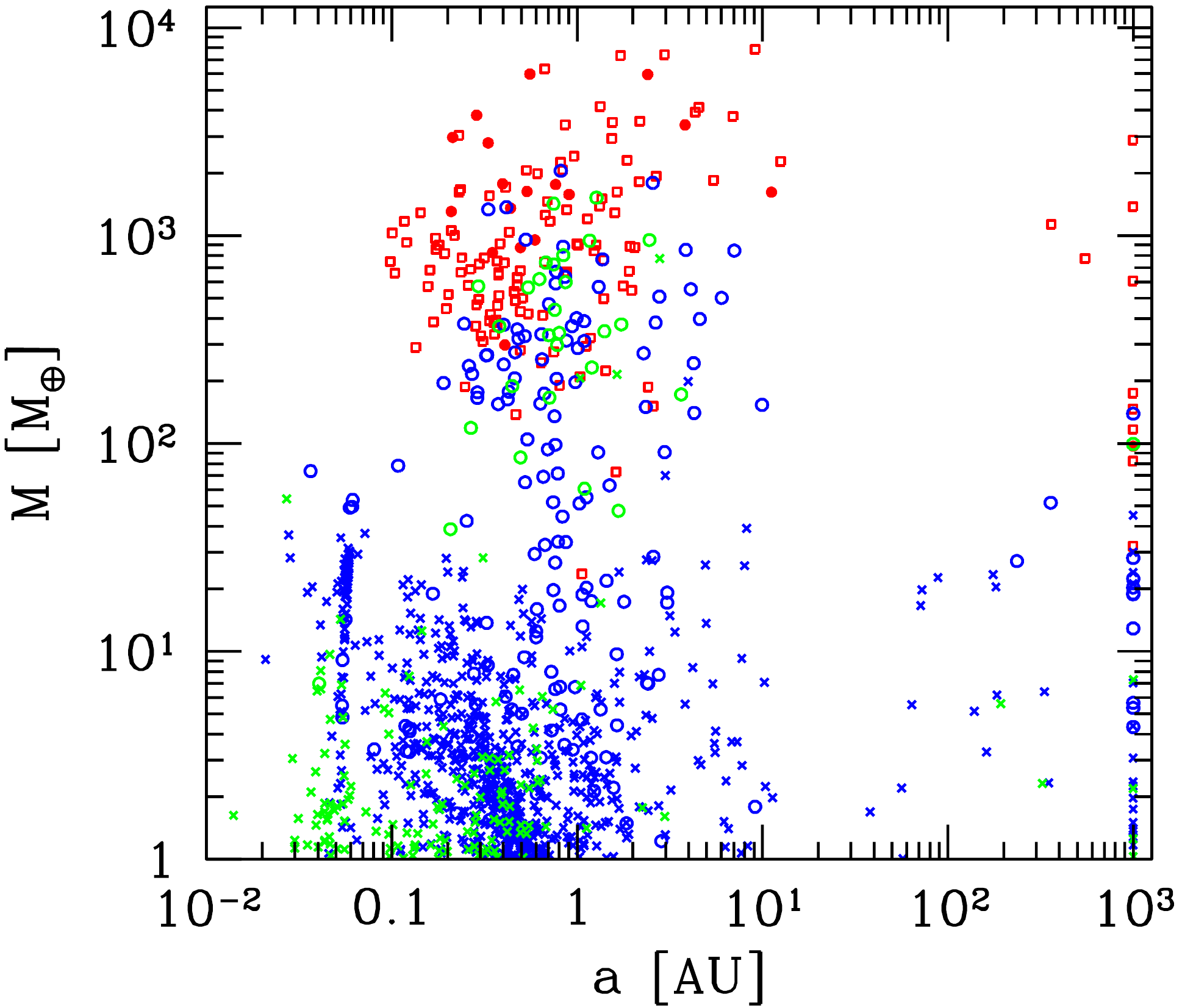}
 \caption{\small Synthetic mass-distance diagram at the time the proto-planetary nebula vanishes. The green (blue) crosses are rocky (icy) planets with a gaseous envelope less massive than the core. The open green (blue) circles are rocky (icy) planets with an envelope 1-10 times more massive than the core. The red filled circles (empty squares) are giant planets with a rocky (icy) core. The envelope is at least 10 times more massive than the core. The model assumes ten proto-planets concurrently forming per disc.}   
 \label{fig:aM10embryo}
 \end{figure}

\begin{figure}[htp]
   \epsscale{1.0}       
   \plotone{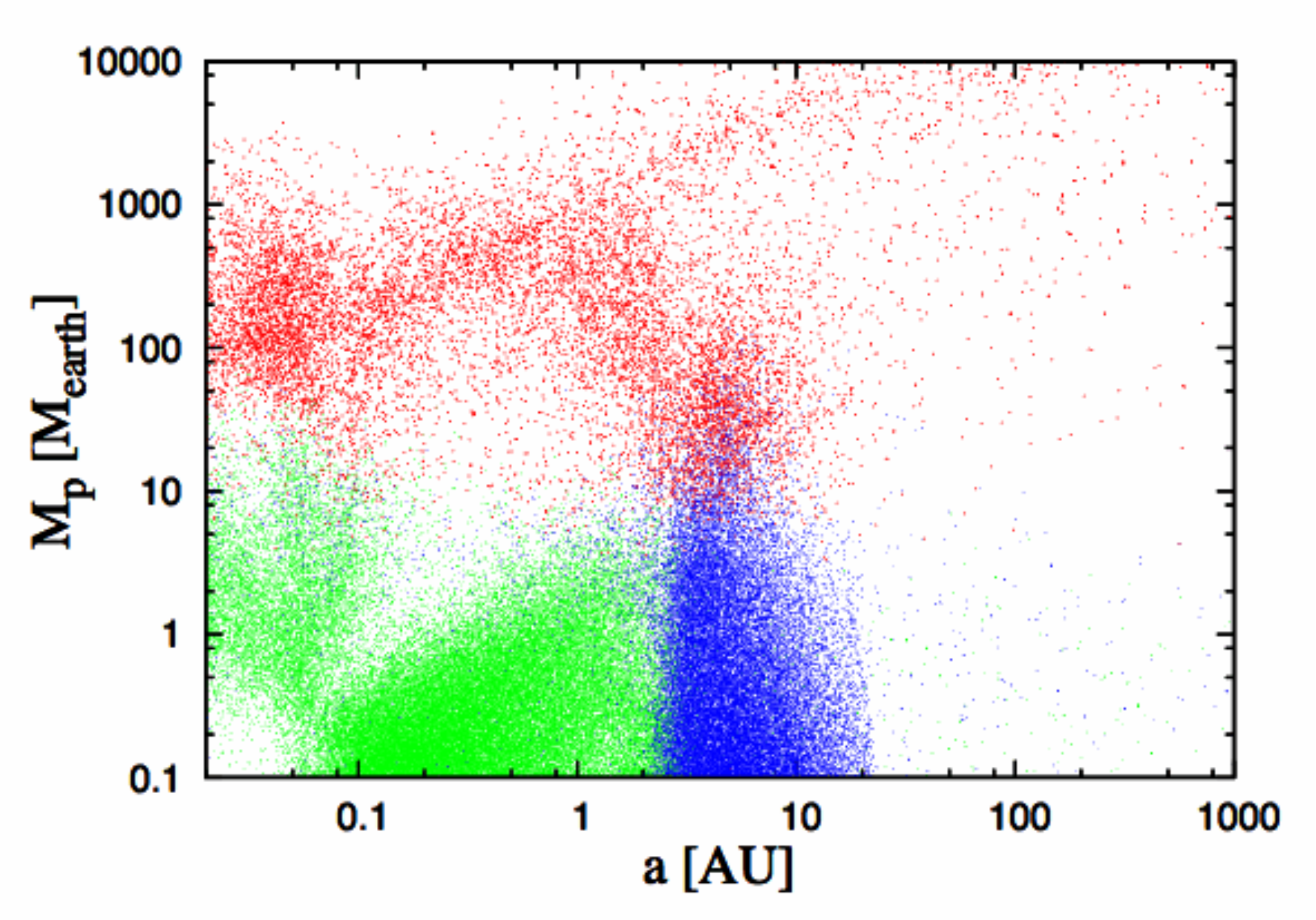} 
     \caption{\small Synthetic mass-distance diagram at the time the proto-planetary nebula vanishes obtained by \citet{idaetal:2013}. The colours have the same meaning as in as Fig.~\ref{fig:aM10embryo}. The population of 0.3-$30M_\oplus$ at $a \la 0.1$AU is formed by in situ merging of proto-planets that have migrated to the vicinity of the host stars
}
  \label{fig:maIL13}
\end{figure}

This is illustrated in Fig.~\ref{fig:aM10embryo} from \citet{alibertetal:2013}. One certainly notes that there are less planets in the 20-100 $\mearth$ mass range than less and/or more massive planets. This part of the diagram is however far from being totally empty.  \citet{idalin:2010} and \citet{idaetal:2013} showed that super-Earths can be formed by {\it in situ} collisional coalescence of two or more proto-planets after disc-gas depletion. Some of these planets have migrated to the vicinity of their host stars before disc-gas depletion, which is illustrated in Fig.~\ref{fig:obt}. This effect, in addition to the limitation of gas accretion onto the planet, also supplies a population of intermediate-mass close-in planets, making the "planetary desert" less conspicuous. Nevertheless, this mass range remains an under-populated region in the $M_p-a$ distribution (see Fig.~\ref{fig:maIL13}).

\begin{figure}[htp]
 \epsscale{1.0}
 \plotone{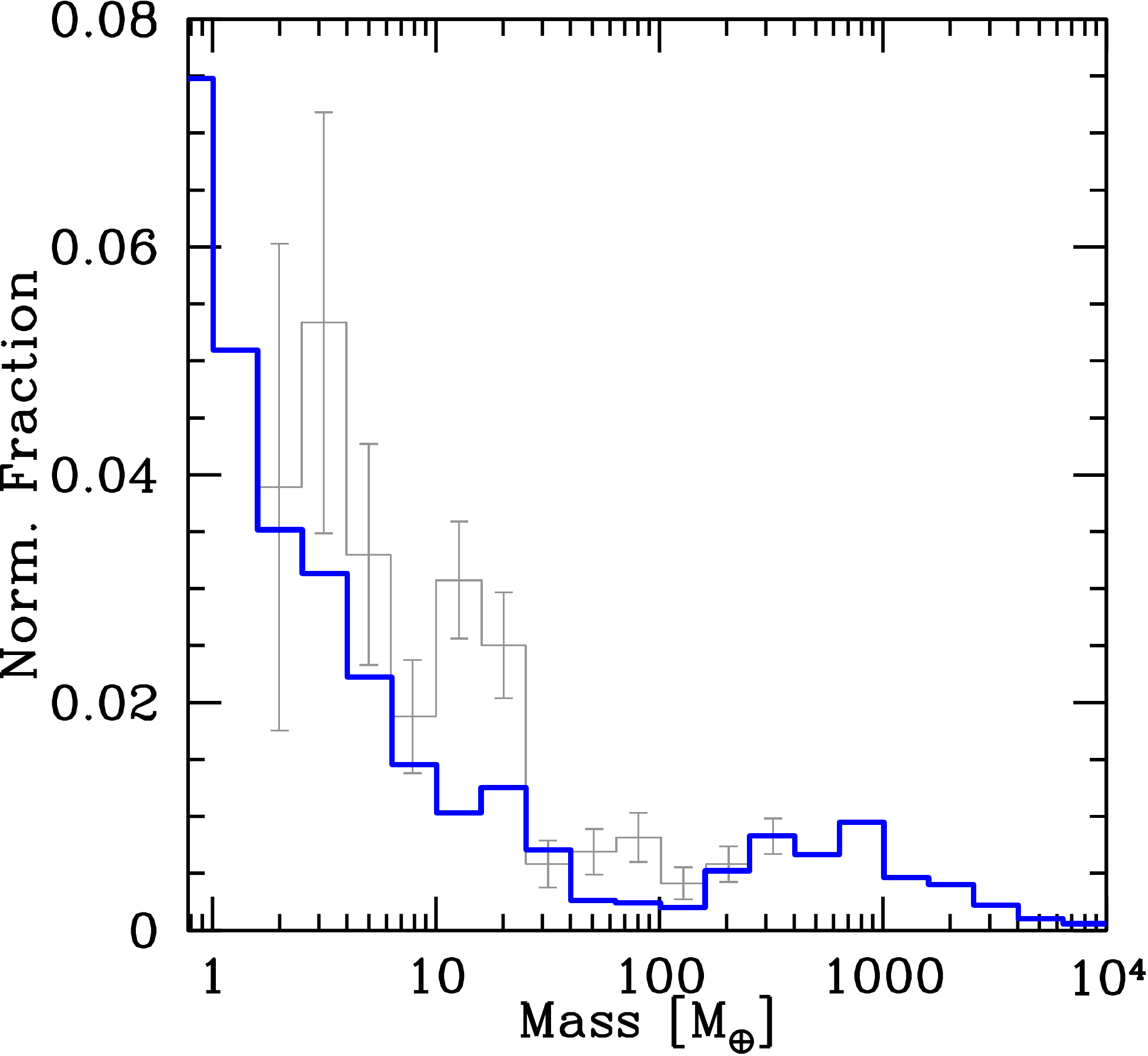}
 \caption{\small Synthetic planetary initial mass function P-IMF (thick  line). The strong increase toward small masses and the transition between solid and gas dominated planets at about 40 $\mearth$ is visible. The thin grey line is the bias-corrected observed mass distribution from  \citet{mayoretal:2011}. It has been normalised to the value of the synthetic distribution in the bin at 1 Jovian mass.}    
 \label{fig:PIMF}
 \end{figure}

A certain depletion of intermediate mass planets is also evident in Fig.~\ref{fig:PIMF}. It shows the planetary initial mass function (P-IMF) \textit{i.e.}, the distribution of planetary masses during the phase when the proto-planetary discs disappear, as found by AMB. The P-IMF is one of the most important results of planetary population synthesis.  Two peaks are clearly present, a smaller one at a few hundred Earth masses representing gaseous giant planets, and a much larger second one for low-mass, solid-dominated planets. It is the simple consequence of the fact that typically, the conditions in the proto-planetary discs are such that only low-mass planets can form. Note that the abundant population of low-mass planets was predicted by population synthesis long before its existence was confirmed by RV surveys and the Kepler mission. The mass function presented in Fig.~\ref{fig:PIMF} is derived from models considering the formation of systems, seeded by 10 proto-planets per disc \citep{alibertetal:2013}. Interestingly enough, the mass function obtained in this case is quite similar to the one obtained with only one proto-planet present in each disc \citep{mordasinietal:2009b}.
 
\begin{figure*}[ht]
   \epsscale{1.6}       
   \plotone{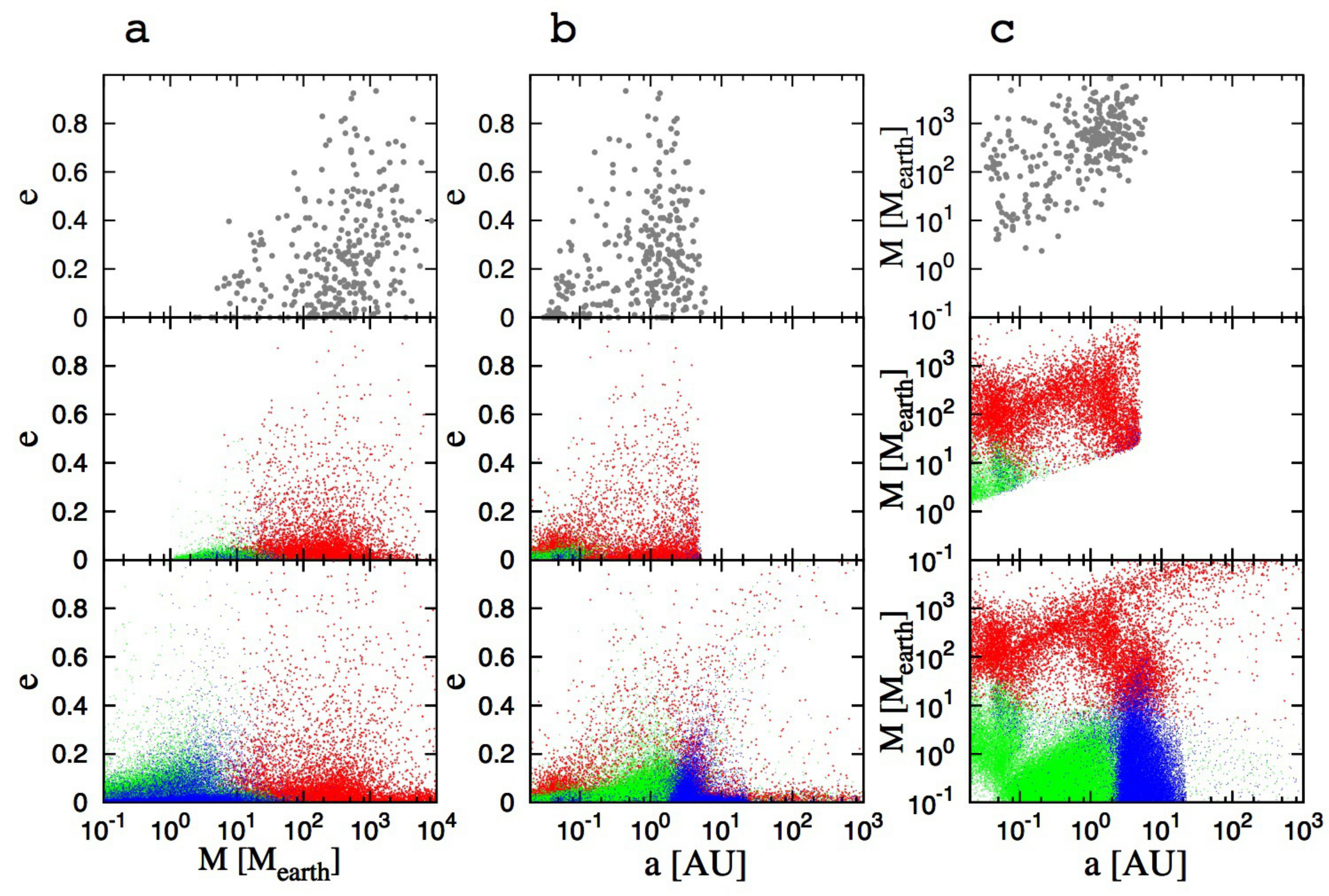} 
      \caption{\small Comparison of theoretical results with RV observational data: (a) $e-M_p$, (b) $e-a$, and (c) $M_p-a$ distributions. The top row are the data obtained from radial velocity surveys. The bottom row are the distributions of planets in $10^4$ systems  simulated by \citet{idaetal:2013} (for detailed simulation parameters, see \citet{idaetal:2013}). The middle row represent only the simulated observable  planets (radial velocities greater than 1 m/s and periods less than 10 years). [from \citealt{idaetal:2013}]
   }
  \label{fig:mea_comp_obs}
\end{figure*}

If we plot all the exoplanets discovered by RV surveys (e.g., http://exoplanets.org/), a deficit of planets in $30-100M_\oplus$ is suggested at $a \la 0.1$AU (see the top panel of Fig.~\ref{fig:mea_comp_obs}c), even though it is not very pronounced.  Results from RV surveys for controlled samples \citep[e.g.][]{howardetal:2010, mayoretal:2011} do not show this deficit. However, their statistical significance is limited by the size of the samples. The Kepler data does not show a clear deficit either, although the distribution function of physical radii found by Kepler can only be translated into a mass distribution if the planetary mass-radius relationship is known, as discussed below \citep[for a comparison of the synthetic planetary radius distribution and  the Kepler results, see][]{mordasinietal:2012b}.  The issue of the "planetary desert" must be further investigated from both observational and theoretical sides.

Interestingly, even if the bias-corrected observed mass distribution of \citet{mayoretal:2011} does not show a deficit at intermediate masses, it nevertheless exhibits a very clear change in the slope of the mass function at about 20-30 $\mearth$  (Fig.~\ref{fig:PIMF}). In the synthetic mass distribution a similar, even though less abrupt change can be seen.  It arises from the fact that when the total mass of the planet is approximately 30 $\mearth$, the accretion of gas becomes very rapid. This is because at this mass, the core already exceeds significantly the critical/crossover core mass which is typically about 15 $\mearth$ \citep{pollacketal:1996}.  The change in slope at about 30-40 $\mearth$ therefore potentially represents the transition from solid to gas dominated planets. In other words, it is evidence of the existence of a critical core mass, a key concept within the core accretion paradigm. If this imprint of core accretion into the planetary mass function is confirmed, it would represent a key statistical finding for both theory and observation.

The observed $M_p-a$ distribution (the top panel of Fig.~\ref{fig:mea_comp_obs}c) also shows a pile-up of gas giants at $\la 0.05$AU (hot Jupiters) and at $\ga 1$AU (cool Jupiters);  the latter being particularly well pronounced. The theoretical prediction by \citet{idaetal:2013} at  the bottom panel of Fig.~\ref{fig:mea_comp_obs}c show too many hot Jupiters and no clear pile-up of cool Jupiters. The inconsistency may be due to too efficient type II migration that IL adopted in their prescriptions as discussed in section \ref{subsec:gap}. In comparison, the $M_p-a$ distribution obtained by \citet{alibertetal:2013}  (Fig.~\ref{fig:aM10embryo}) shows too few hot Jupiters, which may be due to the under-efficient prescription for type II migration they adopted (see section \ref{subsec:gap}). Their result does not show a clear pile up of cool Jupiters, either. Therefore, the observed $M_p-a$ distribution shows that the prescription as currently implemented does not completely capture all aspects of the problem. 
 
 One of the most important aspects discovered by observations is that many gas giants have large orbital eccentricity, sometimes up to $e \sim 0.9$. This is in contrast to Jupiter and Saturn in our Solar system which have low eccentricities of order $\sim 0.05$. Gravitational scattering between gas giants is one of the most plausible mechanisms that can produce high eccentricities. Many N-body simulations \citep[e.g.][]{marzariweidenschilling:2000,fordetal:2000, zhouetal:2007, jurictremaine:2008, chatterjeeetal:2008, fordrasio:2008} of scattering of two or three gas giants have been carried out and they show a functional form of the eccentricity distribution that is consistent with observations. 
 
 However, initial conditions for these simulations have been artificially specified and did not result from system formation calculations. Therefore, it remains to be seen if systems of giant planets compact enough to result in such scattering can be formed.  Such information can only be obtained from population synthesis simulations that follow the formation of truly interacting systems of planets.  \citep{alibertetal:2013, idaetal:2013}.

In Fig.~\ref{fig:mea_comp_obs}a and c,  $e-M_p$ and $e-a$ distributions predicted by  \citet{idaetal:2013} are compared with observed data. In the observed $e-a$ distribution, close-in planets with $a \la 0.1$AU generally have less eccentric orbits than those with $a \sim 1$AU.  This correlation has been attributed to the orbital circularisation  of close-in planets  \citep[e.g.][]{rasioford:1996, dobbsdixonetal:2004, jacksonetal:2008, matsumuraetal:2010, nagasawaida:2011}. Although tidal effects have not been implemented, the $e-a$ correlation is well established in the results obtained. The maximum eccentricity excited by close scattering between gas giants is $\sim v_{\rm esc}/v_{\rm K}$ \citep{safronovzvjagina:1969}. Since surface escape velocity of the giants  $v_{\rm esc}$ is independent of $a$ whereas the Kepler  velocity ($v_{\rm K}$) is $\propto a^{-1/2}$, $e$ of giants resulted by scattering should be $\propto a^{1/2}$. Thus, we obtain the maximum eccentricity  at $a \sim 0.1$AU is 3 times smaller than that at  $a \sim 1$AU in the simulated models, which well reproduces the observed $e-a$ correlation \citep{idaetal:2013}.

The actually observed $e-M_p$ distribution shows that $e$ increases with mass $M_p$. This correlation which is totally counterintuitive is also reproduced by the simulations. These show that multiple massive giants are preferentially formed in relatively massive discs and these systems are more prone to dynamical instabilities, orbit crossing, and excitation of high eccentricities. This trend is actually responsible for this correlation between $e$ and $M_p$ \citep{idaetal:2013}. This correlation was also suggested by \citet{thommesetal:2008} and N-body simulations of giants with various masses \citep{raymondetal:2010}, although their initial conditions were somewhat artificial.

Population synthesis models also generate a population of massive gas giants ($M_p \ga 10M_J$) with large semi-major axes  ($a \ga 30$AU) (Figs.~ \ref{fig:aM10embryo}, \ref{fig:maIL13}, bottom panels of Figs.~\ref{fig:mea_comp_obs}b and c), which could correspond to planets discovered by direct-imaging.  In these results, the fraction of stars that host such planets is limited to a few percent and most of the planets have low eccentricity ($e \la 0.1$). If they are formed by scattering between gas giants, they should have large $e$. Close inspection for the results shows an alternative path for the formation of distant gas giant planets.  In systems which contain  a gas giant(s), the rapid gas accretion of the first generation of gas giant(s) destabilises the orbits of nearby residual proto-planets and some proto-planets are scattered to large distances.  Since planetesimal accretion rate is low at the large distance, some of the proto-planets start to accrete gas efficiently. The scattered proto-planets initially have high eccentric orbits and they take longer time to pass through their apoastrons, so that they tend to accrete gas from that region that has relatively large specific angular momentum. As a result, their orbits become circularised with a radius comparable to their apoastron radius. This path was already found by E. Thommes (2010,  private communication) through a hybrid N-body and 2D hydrodynamical simulation. Because even a single gas giant can scatter multiple proto-planets outward, we also found systems with multiple distant gas giants in nearly circular orbits.

Another important result of population synthesis models has been the quantitative explanation of the so-called ``metallicity effect'' for gas giant planets. It is well known since more than one decade that the probability to observe a giant planet orbiting a given star increases with the metallicity of the latter \citep[e.g.][]{gonzalez:1997,santosetal:2001,fischervalenti:2005}. Models by \citet{idalin:2005} and later on by \citet{mordasinietal:2009b,mordasinietal:2012c} have quantitatively demonstrated that this metallicity effect is a natural outcome of the core-accretion model. Indeed, in this model, the formation of a gas giant follows the initial building of a planetary core of $\sim 10 \mearth$  (the critical mass; see Eq.~\ref{eq:crit_core_mass}).  Although the critical core mass becomes smaller if the core has accreted most of planetesimals in its feeding zone leading to a smaller rate of planetesimal accretion, envelope contraction occurs within a timescale of Myrs only if the core mass is larger than several $M_\oplus$ (Eq.~\ref{eq:tau_KH}). Such relatively massive cores are more easily formed in metal-rich discs, as explained below.

\begin{figure}[ht]
 \epsscale{1.0}
 \plotone{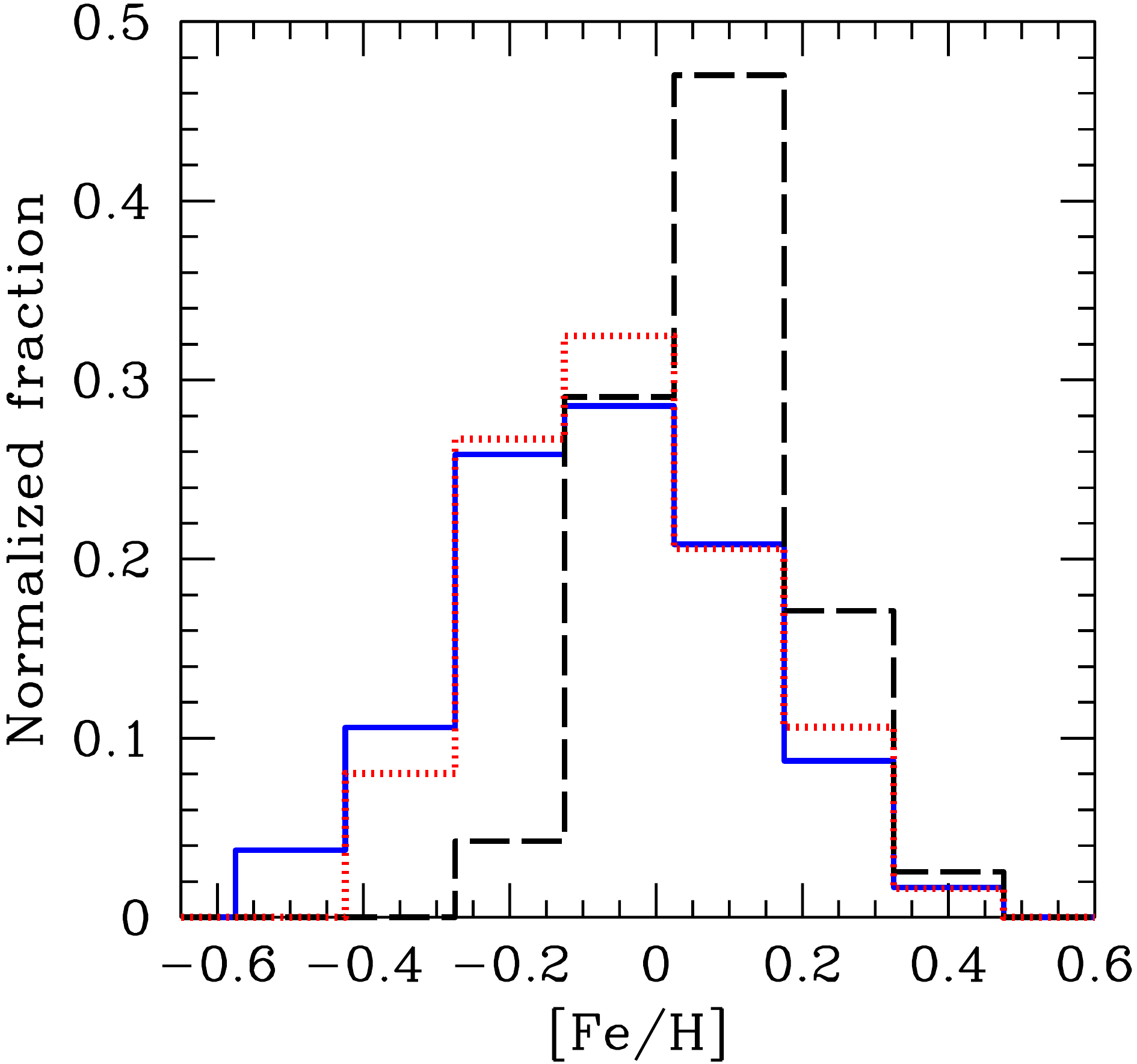}
 \caption{\small Distribution of stellar metallicities in the synthetic population of \citet{alibertetal:2013}. The solid line shows the hosts of all synthetic planetary systems. The dashed line are stars with at least one giant planet ($M\geq100\mearth$) inside of 1 AU, while the dotted line are stars with at least one low-mass planet with  $1\leq M/\mearth\leq 10$. This plot can be compared to Fig. 16 in \citet{mayoretal:2011}. } 
 \label{fig:pfeh}
 \end{figure}

Around a metal rich star, the total solid mass is larger for the same total disc mass as long as metallicity in star and disc are proportional. Then, the building of a planetary core by accretion of planetesimals is faster and the core's isolation mass is larger  \citep[e.g.,][]{kokuboida:1998}, assuming that the amount of planetesimals increases with the mass fraction of heavy elements as indicated by planetesimal formation models \citep[e.g.][]{braueretal:2008}. Planets growing in such environments have a larger likelihood to reach the critical mass before the gas disc has vanished, and are therefore more prone to become observable giant planets. Fig.~\ref{fig:pfeh} presents the distribution of the metallicity of stars harbouring a synthetic planet in a certain mass range as found by AMB in the population presented in \citet{alibertetal:2013}. As initial condition, the total population of stars is assumed to have a metallicity distribution that follows the observed distribution in the solar neighbourhood as described in section \ref{sec:ic} (solid curve). It is clear that the [Fe/H] distribution of stars around which at least one synthetic giant planet forms (dashed curve) is shifted towards higher metallicities, which is the manifestation of the aforementioned ``metallicity effect''. On the other hand, the distribution of stars with Earth to Super-Earth planets (dotted line) is very similar to the one of all stars, showing that there is no metallicity effect for this kind of planets as observed  by radial velocity surveys \citep{sousaetal:2011,mayoretal:2011}.

\citet{matsumuraetal:2013} showed that in systems with multiple gas giants, their secular perturbations often destabilise the orbits of Earths/super-Earths even if  the Earths/super-Earths are in innermost regions far from the gas giants. \citet{idaetal:2013} showed that in discs with larger amount of solid, gas giants often become dynamically active and only a few gas  giants survive in the systems. These results suggest that survival of Earths/super-Earths is inhibited in metal-rich discs, which is consistent with the observation \citep{mayoretal:2011}. In other words, co-existence of Earths/super-Earths and gas giants  may be relatively rare.
 
\subsection{Populations synthesis as an exploratory tool}
%---------------------------------------------------------------------------

\noindent  Even though the current models of population synthesis do not account for all observed characteristics of the discovered exoplanet population, they can nevertheless be used as an exploratory tool for:  1) inferring the combined effects of different processes, and identifying which processes are dominant in shaping the population of extrasolar planets, and 2) inferring the differential effects of different processes  for populations of planets and planetary systems.  To carry out this exploration, two sets of similar models are run by changing only one parameter (or by including/neglecting one process) at a time. We present here a few examples that illustrate the unique power of such an approach.

\subsubsection{Type I migration}
\noindent Since the publication of the first models of migration \citep{goldreichtremaine:1980}, it has been recognised that the very high migration rate derived from linear studies were hardly compatible with the existence of planets. In fact, first models of type I migration predicted a very short migration timescale of $\sim 10^4$ years for a core of $\sim 10 M_\oplus$ at $a\sim5$AU, a time actually much shorter than the growth timescale of the planetary core itself. More detailed later models predicted smaller migration rates \citep[e.g.][]{tanakaetal:2002}, of the order of the accretion timescale of planetesimals. However, the predicted migration timescale still remained much shorter than the observationally inferred typical disc lifetime of a few Myrs. The question arose of the possibility for planets to form and survive given such short migration rates. 

Integrated population synthesis models, including the effects of type I migration, have shown that the formation and survival of planets was indeed possible even with these short migration rates \citep[e.g.][]{idalin:2004a, alibertetal:2005a}. However, the statistics of the planet population obtained with these migration rates \citet{tanakaetal:2002} was quite different from that of the known  planets. These studies also showed that the migration rate had to be reduced by a factor of ${\sim 10^{-1-3}}$ in order to obtain a good agreement between models and observations \citep{idalin:2008a, mordasinietal:2009b, idaetal:2013}. Interestingly enough, the same migration rates (from \citealt{tanakaetal:2002}, reduced by a factor $10-1000$) was also required to match the internal structure and composition of Jupiter and Saturn, using integrated planet formation models \citep{alibertetal:2005b}. 

Population synthesis models have therefore demonstrated that there were some missing physical effects in the original prescription of type I  migration rates, and that  these effects should lead to a global reduction on the overall extent of inward migration of low-mass planets. These findings have helped motivating many investigations on planet-disc interactions.  The missing physical effects were subsequently found to be related to the corotation torque in non-isothermal discs (see the chapter by Baruteau et al. in this book). 

The corotation torque can lead to outward migration provided non-isothermal effects are included. Then, the boundary between outward and inward migration regions, the so-called ``convergence zones'', can act as a migration trap in which high mass planet formation is enhanced \citep[e.g.][]{lyraetal:2010, mordasinietal:2011, kretkelin:2012}. Other possible ``migration traps" are the inner edge of the disc or dead zone \citep[e.g.][]{massetetal:2006, ogiharaetal:2010},  ice line \citep[e.g.][]{kretkelin:2007}, and the outer edge of dead zone \citep[e.g.][]{matsumuraetal:2009, hasegawapudritz:2012, regalyetal:2013}. In order to understand how these migration traps are reflected by distributions of final planets, tests by population synthesis simulations can be useful.

\subsubsection{Type II migration}
%-------------------------------------------
\noindent As discussed in section \ref{subsec:migration}, the description for type II migration rate should be somewhat less uncertain than the one for type I migration. More detailed prescriptions for the transition from type I to type II migration and from disc to planet-dominated migration are nevertheless needed. However, while there is close agreement on these aspects, different assumptions regarding gas accretion across the gap can lead to significant differences in the end distribution of gas giants. 

After the gap is formed, disc gas accretion rate $F$ can be  divided in this region into three components: $f_{\rm mig} F$ the component that pushes the planet (and the inner disc), $f_{\rm p} F$ the fraction that is accreted by the planet, and $f_{\rm cross} F$ the fraction that crosses the gap without being accreted by the planet. In an equilibrium, $f_{\rm mig} + f_{\rm p} + f_{\rm cross} = 1$ ($0 < f_{\rm mig}, f_{\rm p}, f_{\rm cross} < 1$), and the type II migration timescale in the planet-dominated regime is given by $\sim M_p/(f_{\rm mig} F)$ \citep{hasegawaida:2013}. The current version of both IL's and AMB's prescriptions may be over-simplified because they do not take into account the mass flow across the gap. 

For further progress, a prescription for $f_{\rm mig}$, $f_{\rm p}$, and $f_{\rm cross}$ as functions of the planet mass is needed. Several hydrodynamical simulations and analysis have studied these components \citep[e.g.][]{dangeloetal:2002, verasarmitage:2004, lubowdangelo:2006, dobbs-dixonetal:2007, alexanderarmitage:2007, alexanderpascucci:2012}. However, the three components have not been consistently given as functions of planet mass and disc parameters.

\subsubsection{The effect of multiplicity}
%----------------------------------------------------

\noindent By computing two populations based on identical models but with one parameter changed at any one time, it is possible to  assess  the differential effect of some particular processes. To illustrate this, we consider the effect of the number of planets that grow within a given disc. Ultimately, this allows us to compare the formation of a single planet to the formation of a planetary system.

Models by \citet{idalin:2010} and \citet{idaetal:2013} have been the first to consider the emergence of multiple-planet systems in their population synthesis studies. Using a novel approach, that avoid the use of an $N$-body integrator to compute the orbital evolution of the system, they modelled the formation of systems starting with a very large number of growing seeds that grow, migrate and potentially collide with each other. Their approach allows to explore very efficiently the parameter space (since the computation of a population is very rapid), at the expense of neglecting some subtle effects (e.g. low order mean-motion resonances).

Models by \citet{alibertetal:2013} are based on a different approach as they rely on standard $N$-body simulations. While this approach captures virtually all the potential dynamical effects, it comes at a significant computing cost. It is interesting to note that the two approaches are highly complementary, the first approximation allows us to explore extensively the parameter space, and selecting the most interesting cases, whereas the second method can be used to study these interesting cases in more details.

Both studies have shown that 1) the effect of multiplicity is important for low mass planets, and 2) the characteristics of gas giant planets are less affected by the presence multiple planets in a system (with the exception in the context of multiple long-period distant gas giants and dynamical interactions in more closely-packed systems). 

The effects on low-mass planets are: 1) close scattering (eccentricity excitation or ejection) and resonant trapping by large planets and 2) merger events between low-mass planets. The above discussions indicate that close scattering by giant planets  is very important for orbital evolution and survival of low-mass planets in systems with gas giants, in particular, in dynamically active systems.

The oligarchic growth model \citep{kokuboida:1998, kokuboida:2002} predicts that in MMSN, the isolation mass of proto-planets is $\sim 0.1-0.2M_\oplus$ at $\sim 1$AU. While disc gas is present, planet-disc interactions  suppress eccentricity of these proto-planets sufficiently to avoid orbit crossing. After disc gas depletion, they begin to undergo orbit crossing and collisional coalescence among themselves. N-body simulations showed that Earth-sized bodies are formed after multiple collisions \citep[e.g.][]{chamberswetherill:1998, agnoretal:1999, kokuboetal:2006}. This implies that merging of low mass planets after disc gas depletion can increase planetary masses by an order of magnitude. In earlier calculations of IL, this effect was partially included by the artificial expansion of feeding zones after disc gas depletion \citep{idalin:2004a}. But, this merging tendency is consistently included in their recent calculations which directly take into account planet-planet interactions. As stated above, closely-packed multiple super-Earths can be formed near the disc inner edge in a similar way \citep{idalin:2010}. Even in the presence of disc gas, merging between low mass planets occurs when feeding zones overlap due to growth of proto-planets \citep{chambers:2006} or convergent migration between proto-planets occurs \citep{idaetal:2013} . These effects enhance the formation of sufficiently massive cores to initiate runaway gas accretion. 

\subsubsection{Dependence on the host stars' mass}
%--------------------------------------------------------

\noindent Models by \citet{idalin:2005} and \citet{alibertetal:2011} have shown that the population of planets depends upon the mass of the central star. Because disc mass is generally smaller around less massive stars, the frequency of cores reaching a sufficiently large mass to evolve into gas giants is lower in theses systems. This is a natural consequence of the core accretion model. In contrast,the disc instability model does not necessarily imply such a trend. Thus, the correlation between the mass of the host stars and the fraction of stars with gas giants is a good test discriminating between core accretion and disc instability models (for detailed discussions, see \citealt{idalin:2005}).

 \citet{laughlinetal:2004}, \citet{idalin:2005} predicted that gas giants are rare around M stars, while Neptunes are rather abundant. This trend was confirmed later by RV surveys, although micro-lensing surveys  suggests the possibility of an abundant population of gas giants around M dwarfs \citep{gouldetal:2010}. Around Intermediate-mass stars, the fraction of stars with gas giants also increases with mass because there are more heavy elements in their discs to provide building blocks for cores of gas giant planets. Radial velocity surveys suggest that $\eta_{\rm J}$ increases with the stellar mass for $M_\ast$ up to $\sim 1.8 M_\odot$ \citep{johnsonetal:2010}. These planets and their progenitor cores form preferentially beyond the ice line.  Around massive stars with high stellar luminosity, the gas depletion process is rapid and the ice line is located at large disc radii where the growth time for sufficiently massive cores is long.  It has been suggested that the frequency of gas giants ($\eta_{\rm J}$) may reach a maximum for some critical stellar mass and then decreases above this value\citet{idalin:2005, kennedyetal:2007}. Further observational determination of the $\eta_{\rm J}-M_\ast$ correlation in the high stellar mass limit will provide an important constraint for the theoretical models. 

Recent RV surveys for GK clump giant stars (stars that were A dwarfs during main sequence phase) show that there may be a lack of  giant planets inside 0.6 AU for stars with mass $\ga 1.5M_\odot$ (\citealt{satoetal:2010}  and references therein). \citet{kunitomoetal:2011} suggested that tidal decay may not be  responsible for the depletion for $M_* \ga 2M_\odot$. A possible deficit of intermediate period gas giants around F dwarfs were addressed by population synthesis, taking into account possible dependence of disc structure \citep{kretkeetal:2009} or lifetime on stellar mass \citep{burkertida:2007, currie:2009, alibertetal:2011}. The period distribution of gas  giants around A and F dwarfs is another important issues that should be addressed by population synthesis simulation.

The characteristics of planetary systems is sensitively affected by dependence of disc dynamical/thermal structure and its evolution on stellar mass. These disc properties will be revealed by observations by ALMA. Population synthesis is a best tool to test effects of these disc properties on planet formation \citep[see][]{idalin:2005, mordasinietal:2012c} for systematic studies how disc properties translate into planetary properties.

\subsubsection{The planetary mass-radius relationship}
%--------------------------------------------------------------------------

\noindent In the last few years, the study of exoplanets has gone beyond the discovery of data points in the mass-distance diagram. Thanks to various observational techniques complementary to the radial velocity surveys, namely  transit, direct imaging, and spectroscopic observations, it has become possible to derive a characterisation of exoplanets in terms of their basic physical properties like mean density, intrinsic luminosity, and atmospheric composition. In particular the planetary mass-radius relationship has emerged as a new observational constraint for formation theory, since it allows the fundamental geophysical classification of planets (rocky, ice-dominated, gas-dominated).

The population syntheses of  IL and AMB predict not only the total mass of synthetic planets, but also their bulk composition (see Figs.~\ref{fig:aM10embryo} and \ref{fig:maIL13}). This bulk composition acquired during formation determines (for a given orbital distance and entropy for gas-dominated planets) the planetary radius (neglecting special evolutionary effects such as envelope heating and inflation for close-in planets). The planets' mass-radius distribution adds new constraints for population synthesis models.

As an illustration, we first consider the bulk composition of close-in, low-mass planets. These planets have been found in large numbers both by radial velocity surveys and by the Kepler mission. If these planets form inside the iceline, either approximately in situ \citep[e.g.][]{chianglaughlin:2013} or with migration only inside of  $a_{\rm ice}$ (as in Fig.~\ref{fig:obt}), they would have rocky cores.  Alternatively, if these planets (or their building blocks) form outside the iceline, and then migrated inwards over large distances, they would mostly contain ices. If it is  observationally possible to (statistically) distinguish these compositions, it would add strong constraints on type I migration models. Some complications may arise from the (partial) degeneracy of the mass-radius relationship \citep[e.g.,][]{valenciaetal:2007}. A second example is the abundant low-mass, low-density planets \citep[e.g., around Kepler-11][]{lissaueretal:2013}. These low-mass planets seem to have accreted significant amounts of H/He. The efficiency at which a core can accrete gas during the nebular phase depends on the opacity due to grains in the proto-planetary atmosphere.  Theoretical grain evolution models \citep{podolak:2003,movshovitzpodolak:2008} predict low grain opacities, allowing low-mass cores to accrete much H/He, resulting in large radii.

\begin{figure}[ht]
 \epsscale{1.0}
 \plotone{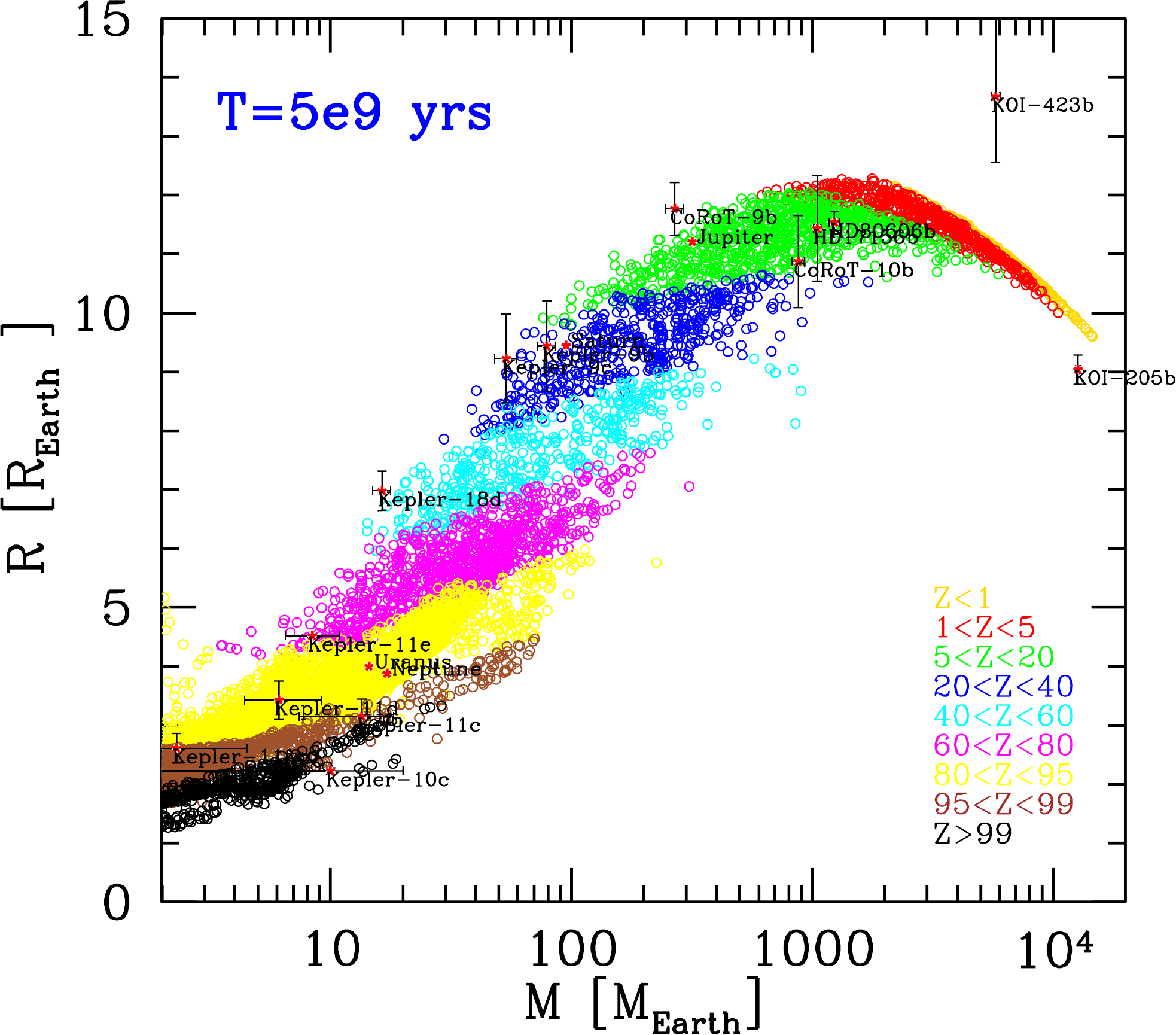}
 \caption{\small Synthetic mass-radius relationship of planets with primordial H/He envelopes at an age of 5 Gyrs compared to Solar System planets and exoplanets with a semi-major axis of at least 0.1 AU. The mass fraction of heavy elements Z (in percent) is given by the scale shown in the plot \citep[updated from][]{mordasinietal:2012b}.}
 \label{fig:MR}
 \end{figure}
 
 In order to take advantage of these new information, AMB expanded their original formation model \citep{alibertetal:2005a} into a self-consistently coupled formation and evolution model (see \citet{mordasinietal:2012} for details) that predicts on a population level planetary radii (and luminosities). Fig.~\ref{fig:MR} from \citet{mordasinietal:2012b} shows the mass-radius relationship for synthetic planets around solar-like stars at an age of 5 Gyrs. The mass-radius distribution has a typical S-shape, with large (in term of radius) planets absent for small masses, and small planets absent in the high-mass domain. It is also clear that there are correlations between the composition and radius and between composition and mass (gas-dominated planets do not exist in the low-mass domain). These correlations are natural consequence of the core accretion mechanism (and of the EOS of different materials): low-mass cores have long KH timescales for envelope accretion (Eq. \ref{eq:tau_KH}), therefore they remain solid dominated, and small, leaving the upper left part of the figure empty. On the other hand, the lower right part remains empty because massive, supercritical cores must accreted massive H/He envelopes (at least if they form during the presence of the nebula), so that their radius is large.  As can be seen in Fig.~\ref{fig:MR}, all the observations can be relatively well reproduced by theoretical models. This good match depends however on the assumed grain opacity in planetary envelopes during their formation. By comparing the observed mass-radius relationship with the synthetic models obtained with different grain opacities, it becomes possible to observationally constrain this important quantity.

\bigskip
\section{CONCLUSIONS AND OUTLOOK}
\bigskip

As demonstrated in this chapter, planet population synthesis is based on a physical description of a large number of processes playing a role in the formation of planets. By necessity, this description is simplified in order to keep the problem tractable and to allow the simulation of the formation of planets for a large set of initial conditions. We have presented the two different approaches that have been most used in the literature. These two sets of prescriptions differ essentially by the degree of simplification being adopted and/or the use of fitting formulas stemming from more complete and detailed calculations of individual processes. While differing in these aspects, both approaches aim at developing a self-consistent model in which the relevant processes operate on their proper timescale. This is necessary to account for the numerous feedbacks occurring between the various processes (see Fig.~\ref{fig:schematic}).  

The detailed physical understanding of the many aspects of planet formation from the early condensation of solids and the formation of planetesimals to the runaway accretion of gas in the late stages of giant planet formation taking place within a time evolving proto-planetary disc is essential. As illustrated in the relevant chapters of this book, some aspects can be studied observationally and some only theoretically essentially by means of large-scale numerical simulations of growing physical accuracy. While all these efforts are essential to our understanding, they are all restricted in either spatial or temporal dimensions. 

The goal of population synthesis is to put these space and/or time snapshots together in order to obtain a full picture and a complete history of planet formation from the initial proto-planetary disc to the observed planetary systems. This is essential for at least two reasons: 1) planet formation is not directly observed, only initial conditions (proto-planetary discs) and end-products (planetary systems) are, the link between the two must be provided by theory at least for now; 2) the diversity of characteristics of the ensemble of exoplanets sets statistical constraints on formation models. 

In the end, population synthesis is similar to idealised models developed by theorists in an attempt to understand the basic behaviour of complex physical systems. The success of the approach does not lie in having all aspects described in detail but to include the key physical processes and their respective feedback. Hence, population synthesis relies on detailed numerical simulations to provide sufficient understanding to allow the derivation of simplified description that capture the essence of the phenomenon. In return, the approach allows visualising the effect on the ensemble population of planets of a given physical description of individual processes (\textit{e.g.} disc structure, migration, opacity) which, when compared to observation, allows setting constraints on the models. 

We have shown in this chapter that this approach has been useful in the past to identify key problems in the theoretical descriptions on key processes active in planet formation (\textit{e.g.} planetary migration). We also have pointed out a number of areas where further detailed modelling is needed. Among those, we highlight the following: 1) disc structure and evolution and the associated transport of gas and solids; 2) the formation of planetesimals and the resulting chemical composition and size distribution as a function of distance to the star;  3) migration is still too fast even taking into account the non-isothermal effects associated with the co-rotation torques; 4) the gas flow through the gap is still not clearly established for the accurate determination of gas giant planets' asymptotic masses. The predictive power of population synthesis will rest on the progress that will be achieved in the future in the physical understanding of these processes.

\bigskip
\noindent\textit{Acknowledgements.}
WB would like to acknowledge partial support from the Swiss National Science Foundation. SI acknowledges support from the JSPS grant. CM acknowledges support from the Max Planck Society through the Reimar-L\"ust Fellowship. YA acknowledge the support of the European Research Council under grant 239605. The authors have been supported by the International Space Science Institute, in the framework of an ISSI Team. DL acknowledge support from NASA, NSF, and UC/Lab fee grants.

\bigskip
%\newpage

\bibliographystyle{ppvi_lim1.bst}
\bibliography{cit}

\end{document}